\documentclass[preprint, 10pt]{elsarticle}
\usepackage[a4paper, total={6.5in, 9in}]{geometry}
\usepackage{amssymb}
\usepackage{amsfonts}
\usepackage{amsmath}
\usepackage{amssymb}
\usepackage{comment}
\usepackage{glossaries}
\usepackage{hyperref}
\glsdisablehyper
\usepackage{mathtools}
\usepackage{pgfplots}
\usepgfplotslibrary{groupplots}
\usepackage{placeins}
\usepackage{siunitx}
\usepackage{subcaption}
\usepackage{tabularx}
\usepackage{tikz}
\usetikzlibrary{shapes.geometric, arrows, patterns}

\journal{Journal}

\biboptions{sort&compress}

\newacronym{anova}{ANOVA}{analysis-of-variance}
\newacronym{cfd}{CFD}{computational fluid dynamics}
\newacronym{cmaes}{CMA-ES}{covariance matrix adaptation - evolution strategy}
\newacronym{ddpce}{DD-PCE}{data-driven polynomial chaos expansion}
\newacronym{de}{DE}{differential evolution}
\newacronym{dl}{DL}{deep learning}
\newacronym{ga}{GA}{genetic algorithm}
\newacronym{gsa}{GSA}{generalized simulated annealing}
\newacronym[firstplural=Gaussian processes (GPs)]{gp}{GP}{Gaussian process}
\newacronym{gpr}{GPR}{Gaussian process regression}
\newacronym{igbt}{IGBT}{insulated gate bipolar transistor}
\newacronym{lar}{LAR}{least angle regression}
\newacronym{ml}{ML}{machine learning}
\newacronym{nn}{NN}{neural network}
\newacronym{nsga}{NSGA}{non-dominated sorting genetic algorithm}
\newacronym{lasso}{LASSO}{least absolute shrinkage and selection operator}
\newacronym{ls}{LS}{least squares}
\newacronym{pdf}{PDF}{probability density function}
\newacronym{pce}{PCE}{polynomial chaos expansion}
\newacronym{pso}{PSO}{particle swarm optimization}
\newacronym[longplural=quantities of interest]{qoi}{QoI}{quantity of interest}
\newacronym{rans}{RANS}{Reynolds-averaged Navier-Stokes}
\newacronym{rms}{RMS}{root-mean-square}
\newacronym{rv}{RV}{random variable}
\newacronym{svm}{SVM}{support vector machine}
\newacronym{svr}{SVR}{support vector regression}
\newacronym{sst}{SST}{shear stress transport}
\newacronym{uq}{UQ}{uncertainty quantification}

\DeclareMathOperator*{\argmin}{arg\,min}

\begin{document}

\begin{frontmatter}

\title{Power Module Heat Sink Design Optimization with Ensembles of Data-Driven Polynomial Chaos Surrogate Models}


\author[1,2]{Dimitrios Loukrezis\corref{cor1}}
\ead{dimitrios.loukrezis@tu-darmstadt.de}
\author[1,2]{Herbert De~Gersem}
\ead{herbert.degersem@tu-darmstadt.de}

\address[1]{Institute for Accelerator Science and Electromagnetic Fields, Technische Universität Darmstadt, Darmstadt, Germany}
\address[2]{Centre for Computational Engineering, Technische Universität Darmstadt, Darmstadt, Germany}

\cortext[cor1]{Corresponding author.}

\begin{abstract}
We consider the problem of optimizing the design  of a heat sink used for cooling an insulated gate bipolar transistor (IGBT) power module.
The thermal behavior of the heat sink is originally estimated using a high-fidelity computational fluid dynamics (CFD) simulation, which renders numerical optimization too computationally demanding.
To enable optimization studies, we substitute the CFD simulation model with an inexpensive polynomial surrogate model that approximates the relation between the device's design features and a relevant thermal quantity of interest. 
The surrogate model of choice is a data-driven polynomial chaos expansion (DD-PCE), which learns the aforementioned relation by means of polynomial regression.
Advantages of the DD-PCE include its applicability in small-data regimes and its easily adaptable model structure.
To address the issue of model-form uncertainty and model robustness in view of limited training and test data, ensembles of DD-PCEs are generated based on data re-shuffling. Then, using the full ensemble of surrogate models, the surrogate-based predictions are accompanied by uncertainty metrics such as mean value and variance.
Once trained and tested in terms of accuracy and robustness, the ensemble of DD-PCE surrogates replaces the high-fidelity simulation model in optimization algorithms aiming to identify heat sink designs that optimize the thermal behavior of the IGBT under geometrical and operational constraints. 
Optimized heat sink designs are obtained for a computational cost much smaller than utilizing the original model in the optimization procedure. 
Due to ensemble modeling, the optimization results can also be assessed in terms of uncertainty and robustness.
Comparisons against alternative surrogate modeling techniques illustrate why the DD-PCE should be preferred in the considered setting. 
\end{abstract}


\end{frontmatter}


\section{Introduction}
\label{sec:intro}
The wide spread of power electronic modules in diverse applications, ranging from common household devices to transportation systems and renewable energy sources, has resulted in high-quality standards in terms of operational reliability and robustness \cite{rashid2010power}.
To meet these demands, design processes of high precision are necessary.
An essential challenge in power electronics design is restraining the temperature of an operating power module below an acceptable limit.
To that end, it is necessary to complement the power module with a cooling system, typically a heat sink, such that the heat generated due to semiconductor power losses is not accumulated within the module.
Given specific operation features, e.g. power losses or ambient environment conditions, the heat sink must be designed such that the thermal stability and, thus, the operational reliability of the power module is guaranteed.
At the same time, industrial demands often require that a suitable design must be delivered within strict time-frames, hence, the need for fast albeit still reliable design methods arises.

A common approach in heat sink design is to combine simulation models with numerical optimization methods in order to optimize the heat sink's geometry with respect to a \gls{qoi} related to the thermal behavior of the device \cite{bornoff2011heat, dede2015topology, maranzana2004design, pakrouh2015numerical, shih2004optimal}. 
This optimization is commonly conducted under cost- and size-related constraints.
While unarguably successful, this design approach faces the bottleneck that the thermal or multi-physics simulations that are necessary to accurately approximate the thermal behavior of the device are usually time- and resource-demanding.
Combined with the need to explore a possibly high-dimensional parameter space comprising operational and geometrical design features and run simulations for multiple configurations until an optimal set of feature values is identified, design optimization tasks often amount to an undesirable computational cost.
This problem is especially relevant if stochastic optimization algorithms are employed \cite{alrasheed2007modified, alrasheed2007evolutionary}, which, despite typically being slower than deterministic optimization methods \cite{ion2018robust}, are a popular choice in the context of engineering design due to their seamless, black-box application, as well as due to their ability to avoid local optima.

One way to accelerate optimization or other resource and time-intensive parametric studies is to replace the computationally demanding simulation with an inexpensive albeit sufficiently accurate surrogate model \cite{forrester2007multi, forrester2008engineering, forrester2009recent, koziel2013surrogate, ong2003evolutionary}. 
Among several surrogate modeling techniques available, an increasingly popular approach is to learn a model using existing design and operation data \cite{cozad2014learning}. 
Owning to the ever improving data collection and storage capacities, many recent works have focused on \gls{ml} methods \cite{huang2008machine}, particularly on \gls{dl} \cite{goodfellow2016deep} and on the use of \glspl{nn} as surrogate models \cite{tripathy2018deep, zhu2019physics}.
However, \gls{dl}-based surrogate modeling typically requires a large amount of available data to train and test the \gls{nn} model.
This can be a crippling problem in cases where databases of sufficient size are not readily available and only a few data points can be generated within an acceptable time-frame.
This is typically the case for numerous complex engineering applications, including heat sink design, e.g. considering data generated with the aforementioned computationally demanding simulations.

Motivated by the need for a data-driven surrogate modeling method which is applicable in small-data regimes, in this work we turn to classical statistical learning theory \cite{vapnik2013nature} and use polynomial regression to derive surrogate models known as \glspl{pce} \cite{ghanem1990polynomial, wiener1938chaos, xiu2002wiener}.
\glspl{pce} were originally developed within the context of \gls{uq}, where they are known to provide accurate and cost-efficient moment and sensitivity estimates for random \glspl{qoi} \cite{crestaux2009polynomial, sudret2008global}.
More recently, \glspl{pce} have been applied within a supervised \gls{ml} context, i.e. for the purpose of providing purely data-driven predictive models \cite{torre2019data}.
The method was named \gls{ddpce} and was found to perform comparably well against state-of-the-art \gls{ml} methods, e.g. based on deep \glspl{nn} and \glspl{svm}.
Besides model learning with small data, the choice of the \gls{ddpce} method is further justified by the fact that, if combined with basis selection algorithms \cite{abraham2017robust, blatman2010adaptive, blatman2011adaptive, cheng2018sparse, diaz2018sparse, doostan2011non, hampton2018basis, he2020adaptive, hu2011adaptive, huan2018compressive, loukrezis2020robust, migliorati2013a, peng2014weighted, tsilifis2019compressive, zhao2019efficient}, the model structure of a \gls{ddpce} can be easily adapted according to the influence of individual design features, desired approximation accuracy, and data availability.
This is a significant advantage compared to \gls{ml} models of fixed structure, e.g. deep \glspl{nn} with a pre-determined architecture.

This work presents a framework based on surrogate modeling with \glspl{ddpce} that enables otherwise intractable optimization studies with respect to the design of an air-cooled heat sink attached to an \gls{igbt}.
Our goal is to identify a heat sink design that results in the desired thermal behavior of the device, given application-dependent operational and geometrical constraints.
The monitored \gls{qoi} is the steady-state temperature of the heat sink, which is directly connected to the maximum temperature developed in the power module during its operation.
For a given design, the heat sink's temperature is originally estimated using a high-fidelity, 3D \gls{cfd} simulation.
Due to the duration of a single simulation, using the 3D model in an optimization algorithm would result in an unacceptable computational cost, especially if a stochastic optimization algorithm is employed.
Instead, we train a \gls{ddpce} to approximate the dependency of the \gls{qoi} on the design features using a relatively small set of simulation-generated data.
A separate dataset is employed to test the predictive capability of the \gls{ddpce} surrogate model.
Note that the training and test data are labeled, i.e. the datasets consist of temperature values for different design feature instances, hence, \gls{ddpce} constitutes a supervised \gls{ml} method. 

However, the use of training and test datasets of relatively small size poses the question of model robustness against variations in the data. Equivalently, the form of the surrogate model might be significantly altered when different data are employed. This leads to so-called model-form uncertainties \cite{der2009aleatory}, which inevitably affect the surrogate's accuracy as well. 
To assess the \gls{ml}-based surrogate model in terms of its robustness against variations in the training and test data, an ensemble modeling approach is additionally utilized \cite{goel2007ensemble, ren2022ensemble}, which allows to compute uncertainty metrics with respect to the predictions of the surrogate. 
The ensemble of \gls{ddpce} surrogate models then replaces the original model in the stochastic optimization algorithm, thus enabling the task of heat sink design optimization, while additionally providing uncertainty estimates for the identified optimal design features.

There exist previous works that have applied \gls{pce}-based surrogate models in the context of heat sink design, e.g., for \gls{uq} \cite{sterr2021uncertainty} and optimization \cite{bodla2013optimization, sarangi2014manifold}. 
In contrast to these works, this paper considers the data-driven variant of the \gls{pce} method, which is based on \gls{ml} regression. 
Instead, the aforementioned works have employed collocation methods to compute the \gls{pce}, e.g. based on interpolation on Smolyak sparse grids \cite{bodla2013optimization, sarangi2014manifold}, which is a fundamentally different approach to the supervised \gls{ml}-based \gls{ddpce} method employed in this paper. 
Moreover, central to this work are the issues of model-form uncertainty and model robustness against variations in the datasets that are used for surrogate model learning and verification, which have not received sufficient attention so far. The ensemble modeling approach suggested and utilized in this paper for that purpose, has not been considered in combination with the \gls{ddpce} method before, at least to the authors' knowledge.

The rest of this paper is organized as follows. 
Section~\ref{sec:design} presents the setup of the device under investigation, consisting of an \gls{igbt} and an air-cooled heat sink, and discusses modeling, simulation, and design optimization aspects.
Note that we consider a real-world \gls{igbt} as a test case, typical applications of which include motor control, uninterruptible power supply, and air-conditioning. 
Nevertheless, the framework suggested in this paper is generally applicable and not restricted to this particular power module.
\glspl{ddpce} are presented in Section~\ref{sec:ddpce}.
Section \ref{sec:ensemble} discusses the ensemble modeling approach that is utilized to assess the robustness of data-driven surrogate models against variations in the training and test data.
Numerical studies illustrating the advantages of \glspl{ddpce} in terms of prediction accuracy and robustness compared to alternative data-driven surrogate modeling approaches, as well as the utilization of \glspl{ddpce} ensembles within stochastic algorithms for single and multi-objective optimization, are presented in Section~\ref{sec:applic-num-res}.
Finally, our conclusions are presented in Section~\ref{sec:concl}.

\section{Power module heat sink design: modeling, simulation, and optimization}
\label{sec:design}

\subsection{Device setup}
\label{subsec:setup}

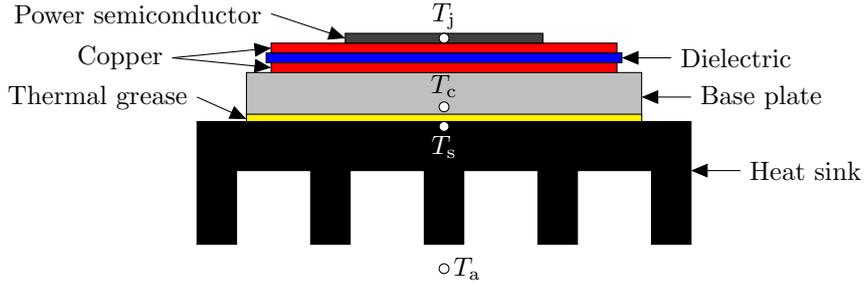
\begin{figure}[t]
	\centering
	\begin{tikzpicture}[scale=0.65]

\draw[fill = black] (-1,0) -- (-1,-2.5) -- (-0.2,-2.5) -- (-0.2,-1) -- (1.3,-1) -- (1.3,-2.5) -- (2.1,-2.5) -- (2.1,-1) -- (3.6,-1) -- (3.6,-2.5) -- (4.4,-2.5) -- (4.4,-1) -- (5.9,-1) -- (5.9,-2.5) -- (6.7,-2.5) -- (6.7,-1) -- (8.2,-1) -- (8.2,-2.5) -- (9,-2.5) -- (9,0) -- cycle;
\draw[fill = lightgray] (0,0.1) rectangle (8,1);
\draw[fill = yellow] (0,0) rectangle (8,0.15);
\draw[fill = red] (0.5,1) rectangle (7.5,1.2);
\draw[fill = blue] (0.4,1.2) rectangle (7.6,1.4);
\draw[fill = red] (0.5,1.4) rectangle (7.5,1.6);
\draw[fill = darkgray] (2, 1.6) rectangle (6,1.8);

\draw[fill = white] (4,1.7) circle (1mm) node[above]{$T_\text{j}$};
\draw[fill = white] (4,0.3) circle (1mm) node[above]{$T_\text{c}$};
\draw[fill=white] (4,-0.1) circle (1mm) node[below, white]{$T_\text{s}$};
\draw[fill = white] (4,-3) circle (1mm) node[right]{$T_\text{a}$};

\draw[triangle 45-] (9,-1) -- (10,-1) node[right] {Heat sink};
\draw[triangle 45-] (0,0.05) -- (-1,0.5) node[left] {Thermal grease};
\draw[triangle 45-] (8,0.5) -- (9,0.5) node[right] {Base plate};
\draw[triangle 45-,] (0.5,1.1) -- (-1.5,1.3) node[left] {Copper};
\draw[triangle 45-] (0.5,1.5) -- (-1.5,1.3);
\draw[triangle 45-] (7.6,1.3) -- (8.6,1.3) node[right] {Dielectric};
\draw[triangle 45-] (2,1.7) -- (0.5,2.2) node[left] {Power semiconductor};


\end{tikzpicture}
	\caption{Illustration of an air-cooled heat sink attached to an IGBT power module, along with common reference temperature positions.}
	\label{fig:structureIGBT}
\end{figure}

An illustration of the \gls{igbt} module and the air-cooled heat sink attached to it is provided in Figure~\ref{fig:structureIGBT}.
The core elements of the \gls{igbt} module are power semiconductors, which generate heat due to conducting and switching losses.
The semiconductors are connected to a direct copper bonded substrate, which acts as an isolation layer and consists of a copper layer, an epoxy-based dielectric layer, and another copper layer.
The lower copper layer is attached to a metallic base plate, which acts as the interface between the power module and the heat sink.
A thin layer of thermal grease is inserted between the base plate and the heat sink to eliminate air gaps in the interface area.
The heat sink consists of a root, which is the part attached to the base plate, thus accumulating heat due to conduction.
This heat is then transferred towards the fins at the bottom of the heat sink, to be exchanged with the air that flows around them.
The heat sink is painted black in order to maximize heat dissipation through radiation.

The heat accumulated within the power module during its operation depends on a number of factors, namely, the power losses $P$ of the semiconductors, the conditions of the surrounding environment, in particular the ambient temperature $T_\text{a}$ and the air flow velocity $v$, and the geometry of the heat sink. 
The latter is described by the length~$l$, the base height~$h_\text{b}$, the fin height~$h_\text{f}$, the number of fins~$N_\text{f}$, the fin width~$w_\text{f}$, and the width of the gap between two fins~$w_\text{g}$. 
All operational and geometrical design features are also shown in Figure~\ref{fig:sim_model}, where a 3D simulation model of the power module and the heat sink is illustrated. 
Since the power module and the heat sink are designed for a certain application range, the design features take values within bounded intervals which are given in Table~\ref{tab:features}.

\begin{table}[b!]
	\centering
	\caption{Geometrical and operational design features and their value ranges.}
	\begin{tabular}{lcccc}
		\hline 
		Parameter & Symbol & Unit & Min. & Max. \\ 
		\hline 
		Heat sink length & $l$ & \si{\milli\meter} & $50$ & $200$ \\ 
		Fin gap & $g_{\text{f}}$ & \si{\milli\meter} & $3$ & $8$ \\ 
		Fin width & $w_{\text{f}}$ & \si{\milli\meter} & $1.4$ & $4$ \\ 
		Fin height & $h_{\text{f}}$ & \si{\milli\meter} & $16$ & $45$ \\ 
		Base height & $h_{\text{b}}$ & \si{\milli\meter} & $4$ & $15$ \\ 
		Number of fins & $N_{\text{f}}$ & -- & $5$ & $25$ \\ 
		Air flow velocity & $v$ & \si{\meter\per\second} & $1$ & $5$ \\ 
		Ambient temperature & $T_{\text{a}}$ & \si{\degreeCelsius} & $25$ & $45$ \\ 
		Power loss & $P$ & \si{\watt} & $115$ & $140$ \\ 
		\hline 
	\end{tabular}
	\label{tab:features}
\end{table}

\subsection{Thermal characterization of power modules}
\label{subsec:qoi}
The thermal management of power electronic devices is typically based on monitoring the thermal impedance $Z_{\text{th}}$ or resistance $R_{\text{th}}$ between two locations of the device, upon which temperature measurements are taken. 
Note that, contrary to electric circuit design, the term ``impedance'' is used when considering transient heat dissipation, whereas the term ``resistance'' refers to stationary, steady-state heat dissipation.
Assuming two locations on the device, respectively denoted with ``1'' and ``2'', the thermal impedance is given by
\begin{equation}
	\label{eq:thermal_impedance}
	Z_{\text{th},{1-2}} = \frac{T_2\left(t\right) - T_1\left(t\right)}{P\left(t\right)}.
\end{equation}
Removing the time dependence, i.e. for steady-state temperatures and power losses, the thermal resistance $R_{\text{th},{1-2}}$ is obtained.

For investigating the thermal behavior of the \gls{igbt} module and the attached heat sink, the most commonly employed temperature positions are shown in Figure~\ref{fig:structureIGBT}. 
These are: the junction temperature $T_{\text{j}}$, measured on the semiconductor chip; the case temperature $T_{\text{c}}$, measured at the bottom of the base plate; the heat sink's temperature $T_{\text{s}}$, measured exactly underneath the contacting surface with the power source, where the maximum temperature developed within the heat sink is observed; and the ambient temperature $T_{\text{a}}$.
The critical measure with respect to the reliable operation of the power module is the junction temperature $T_j$, which must not exceed the threshold value given in the device's specifications.
For the particular \gls{igbt} under investigation, this threshold is $175\si{\degreeCelsius}$.
However, measuring the semiconductors' temperature $T_{\text{j}}$ is out of reach in most physical setups.
Similar measurement difficulties arise for the case temperature $T_{\text{c}}$.
Therefore, most \gls{igbt} specifications include a worst-case estimate for the thermal resistance of the power module $R_{\text{th,j-c}}$, which for the considered test case is $0.74$~\si{\kelvin\per\watt}.
Then, using thermal circuit model theory \cite{gerstenmaier2009combination, ma2015electro, wu2016temperature} and taking into account that the steady-state temperature of the heat sink coincides with its maximum temperature during the operation of the \gls{igbt}, a worst-case estimate for the junction temperature $T_{\text{j}}$ can be obtained using the formula
\begin{equation}
	\label{eq:junction_temp}
	T_{\text{j}} = T_{\text{a}} + \left(R_{\text{th,j-c}} + R_{\text{th,s-a}}\right) P,
\end{equation}
where $R_{\text{th,s-a}}$ refers to the thermal resistance of the heat sink, which is given by
\begin{equation}
	\label{eq:thermal_resistance}
	R_{\text{th,s-a}} = \frac{T_{\text{s}} - T_{\text{a}}}{P}.
\end{equation}
Contrary to the junction and case temperatures, the heat sink's temperature $T_{\text{s}}$ can be easily measured, typically using a sensor drilled under the contact surface.
The ambient temperature $T_{\text{a}}$ is also easy to measure by placing sensors in the surrounding area of the device.
As a result, to ensure the reliable operation of the power module, it suffices to monitor the heat sink's temperature $T_{\text{s}}$.


\subsection{Simulation model}
\label{subsec:simulation}
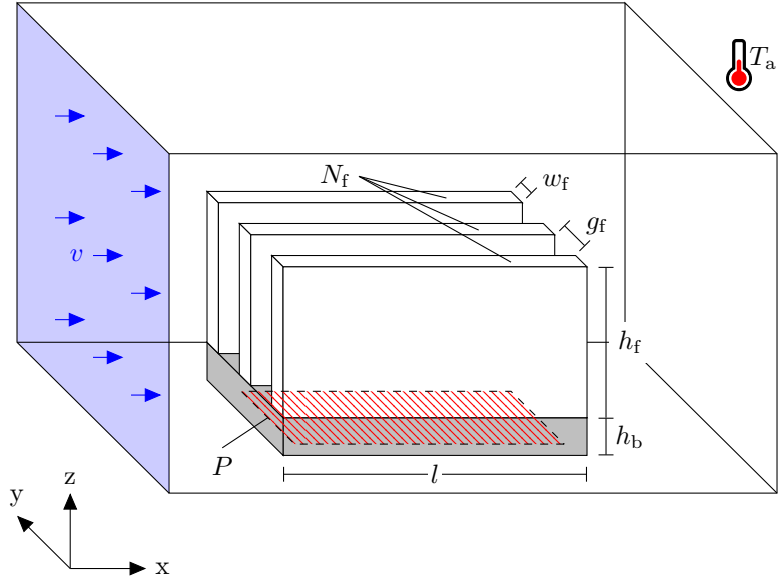
\begin{figure}[t!]
	\centering
	\begin{tikzpicture} 

\draw [blue,-triangle 45] (-9,4.5) -- (-8.6,4.5);
\draw [blue,-triangle 45] (-9.5,5) -- (-9.1,5);
\draw [blue, -triangle 45] (-8.5,4) -- (-8.1,4);

\draw [blue,-triangle 45] (-9,3.15) node[left] {$v$} --  (-8.6,3.15) ;
\draw [blue,-triangle 45] (-9.5,3.65) -- (-9.1,3.65);
\draw [blue, -triangle 45] (-8.5,2.65) -- (-8.1,2.65);

\draw [blue,-triangle 45] (-9,1.8) -- (-8.6,1.8);
\draw [blue,-triangle 45] (-9.5,2.3) -- (-9.1,2.3);
\draw [blue, -triangle 45] (-8.5,1.3) -- (-8.1,1.3);

\draw[very thick](-8+7.5,5+0.5) circle[ radius=0.15];
\draw[rounded corners=1, very thick,fill=white] (-8.07+7.5,5.13+0.4) rectangle node[right]{$T_{\text{a}}$} (-7.93+7.5,5.5+0.5) ;
\fill[fill=white](-8+7.5,5+0.5) circle[ radius=0.13];
\fill[fill=red](-8+7.5,5+0.5) circle[ radius=0.1];
\fill[rounded corners=1,fill=red] (-8.03+7.5,5+0.5) rectangle (-7.97+7.5,5.4+0.35);

\draw (0,0) -- (-8,0) -- (-8,4.5) -- (0,4.5) -- cycle ;
\draw (0,0) -- (-2,2) -- (-2,6.5) -- (0,4.5);
\draw[fill=blue,fill opacity=0.2] (-8,0) -- (-10,2) -- (-10,6.5) -- (-8,4.5);
\draw (-2,2) -- (-10,2);
\draw (-2,6.5) -- (-10,6.5) ;

\draw[fill=lightgray] (-2.5,0.5) -- (-6.5,0.5) --  (-6.5,1) -- (-2.5,1) -- cycle;
\draw[fill=lightgray] (-6.5,0.5) -- (-7.5,1.5) -- (-7.5,2) -- (-6.5,1) -- cycle;
\draw[fill=lightgray] (-2.5,1) -- (-6.5,1) -- (-7.5,2) -- (-3.5,2) -- cycle;

\draw[fill=white] (-7.5,4) -- (-7.5,2) -- (-7.35,1.85) -- (-7.35,3.85) -- cycle;
\draw[fill=white] (-3.5,4) -- (-3.35,3.85) -- (-7.35,3.85) -- (-7.5,4) -- cycle;
\draw[fill=white] (-3.35,3.85) -- (-3.35,1.85) -- (-7.35,1.85) -- (-7.35,3.85) -- cycle;

\draw[fill=white] (-2.925,3.425) -- (-3.075,3.575) -- (-7.075,3.575) -- (-6.925,3.425) -- cycle;
\draw[fill=white] (-6.925,3.425) -- (-6.925,1.425) -- (-7.075,1.575) -- (-7.075,3.575) -- cycle;
\draw[fill=white] (-2.925,3.425) -- (-2.925,1.425) -- (-6.925,1.425) -- (-6.925,3.425) -- cycle;

\draw[fill=white] (-2.5,3) -- (-2.65,3.15) -- (-6.65,3.15) -- (-6.5,3) -- cycle;
\draw[fill=white] (-6.65,3.15) -- (-6.65,1.15) -- (-6.5,1) -- (-6.5,3) -- cycle;
\draw[fill=white] (-2.5,1) -- (-2.5,3) -- (-6.5,3) -- (-6.5,1) -- cycle;

\draw[dashed, pattern=north west lines, pattern color=red] (-2.8,0.65) -- (-6.35,0.65) -- (-7.05,1.35) -- (-3.5,1.35) -- cycle;

\draw[|-|] (-2.25,0.5) -- node[right] {$h_{\text{b}}$} (-2.25,1);
\draw[-|] (-2.25,1) -- node[right,fill=white,xshift=1] {$h_{\text{f}}$} (-2.25,3);
\draw[|-|] (-3.1-0.125,3.85+0.125)  --  (-3.25-0.125,4+0.125) node[right,fill=white, xshift=7, inner sep = 1] {$w_{\text{f}}$};
\draw[|-|] (-2.4-0.125,3.15+0.125) --  (-2.675-0.125,3.425+0.125) node[right,fill=white, xshift=7, inner sep = 1] {$g_{\text{f}}$};
\draw[|-|] (-2.5,0.25) -- node[,fill=white, xshift=0, inner sep = 1] {$l$} (-6.5,0.25);
\draw (-6.7,1) -- (-7.3,0.6) node[below] {$P$};
\draw (-3.35-1,3.85+0.125/2) -- (-5.5,4.2) node[left] {$N_{\text{f}}$};
\draw (-2.925-1,3.425+0.125/2) -- (-5.5,4.2);
\draw (-2.5-1,3+0.125/2) -- (-5.5,4.2);

\draw[triangle 45-] (-8.3,-1) node[right]{x} -- (-9.3,-1);
\draw[-triangle 45] (-9.3,-1) -- (-9.3,0) node[above] {z};
\draw[-triangle 45] (-9.3,-1) -- (-10,-0.3) node[above] {y};

\end{tikzpicture}
	\caption{Illustration of the 3D \gls{cfd} simulation model.}
	\label{fig:sim_model}
\end{figure}

To accurately compute the heat sink's temperature $T_{\text{s}}$ and thus estimate the junction temperature $T_{\text{j}}$ for a given set of operational and geometrical design feature values as described in Section~\ref{subsec:setup}, a 3D \gls{cfd} simulation is used to accurately resolve the heat transfer from the \gls{igbt} to the heat sink and its dissipation to the surrounding environment. 
The simulation model is implemented using the ANSYS Icepak\footnote{https://www.ansys.com/products/electronics/ansys-icepak} software.
Without delving into details, the simulation is based on the finite-volume discretization of the \gls{rans} equations \cite{alfonsi2009reynolds, chen1990solutions, temam2001navier}. 
Additionally, the Boussinesq approximation \cite{gray1976validity} is chosen to model Reynold stresses in the conservation of momentum equation and the $k-\omega$ \gls{sst} model \cite{menter1992influence} is employed for turbulence modeling.
The 3D simulation model is illustrated in Figure~\ref{fig:sim_model} and consists of three main components, namely the \gls{igbt} module, the heat sink, and a cubic box enforcing the necessary boundary conditions. 
In particular:
\begin{itemize}
	\item The \gls{igbt} is not explicitly modeled, but represented only as a rectangular power source of fixed dimensions, shown with red in Figure~\ref{fig:sim_model}. This modeling decision is connected to only considering the worst-case thermal resistance of the power module (see Section~\ref{subsec:qoi}), hence, resolving the heat transfer inside the power module is not necessary.
	Moreover, the power losses are considered to be uniformly distributed on the area of the source. This assumption is justified by the fact that the considered \gls{igbt} module is equipped with a metallic base plate (see Section~\ref{subsec:setup} and Figure~\ref{fig:structureIGBT}).
	\item The heat sink is placed on the power source in order to dissipate the generated heat.
	The material of the heat sink is chosen such that its density and specific heat capacity remain constant over time, while its thermal conductivity is linearly dependent on its temperature.
	\item The cubic boundary box takes the form of an air duct, such that air flows in the positive $x$-direction from the left-side boundary to the right-side boundary. 
	A uniform air flow of constant velocity $v$ and at ambient temperature $T_{\text{a}}$ is imposed on the left-side boundary, while the right-side boundary acts as a pressure outlet.
	The bottom boundary is an adiabatic wall, such that the heat generated  by the power source is fully transferred to the heat sink. 
	The remaining boundaries are open. 
\end{itemize}

\subsection{Heat sink design optimization}
\label{subsec:optimization}

As already mentioned in Section~\ref{subsec:qoi}, the critical measure regarding the reliable operation of the power module is the junction temperature $T_{\text{j}}$, which must not exceed a critical threshold according to the device's specifications. 
Hence, the main design goal is to identify operational and geometrical design features which satisfy this requirement.
In principle, this can be accomplished by solving the optimization problem
\begin{align}
	\label{eq:single-objective-opt}
	\min_{\mathbf{y}} T_{\text{j}}\left(\mathbf{y}\right),
\end{align}
where $\mathbf{y} = \left(l, g_{\text{f}}, w_{\text{f}}, h_{\text{f}}, h_{\text{b}}, N_{\text{f}}, v, T_{\text{a}}, P\right) \in \mathbb{R}^9$ is the design feature vector and $T_{\text{j}}(\mathbf{y})$ denotes the dependency of the junction temperature on the design features.
If the minimum junction temperature $T_{\text{j}}^*$ obtained by solving problem \eqref{eq:single-objective-opt} is below the given threshold, then a suitable heat sink design has been identified.

When solving the optimization problem \eqref{eq:single-objective-opt}, a number of constraints must be additionally taken into consideration.
First, the design features take values within the bounded intervals given in Table~\ref{tab:features}.
Second, the values of the ambient temperature $T_{\text{a}}$ and the power losses $P$ depend on the specific application of the \gls{igbt} and therefore lie outside of the design engineer's control.
On the contrary, the geometry of the heat sink can be fully controlled.
The air flow velocity can also be controlled in most cases, e.g. using a fan.
As a result, a constrained optimization problem must be solved, given as
\begin{align}
	\label{eq:constrained-opt}
	\min_{\mathbf{y}} T_{\text{j}}\left(\mathbf{y}\right) \:
	\text{subject to:} \:\:\mathbf{y} \in \mathcal{D}, P = P_0, T_{\text{a}} = T_{\text{a},0},
\end{align}
where $P_0$, $T_{\text{a},0}$ denote the fixed power losses and ambient temperature, respectively, and $\mathcal{D} \subset \mathbb{R}^9$ is a set containing all admissible realizations of the parameter vector $\mathbf{y}$.


However, solving the constrained optimization problem \eqref{eq:constrained-opt} will typically result in designs which maximize the size of the heat sink, thus also its material requirements and manufacturing cost.
On the contrary, it is desirable that the heat sink remains as compact as possible, while still achieving a sufficiently low junction temperature. 
This requirement leads to the multi-objective optimization \cite{branke2008multiobjective, gunantara2018review} problem
\begin{align}
	\label{eq:multi-objective-opt}
	\min_{\mathbf{y}}\left(T_{\text{j}}(\mathbf{y}), V_{\text{s}}(\mathbf{y})\right) \:
	\text{subject to:} \:\:\mathbf{y} \in \mathcal{D}, P = P_0, T_{\text{a}} = T_{\text{a},0},
\end{align}
where $V_{\text{s}}(\mathbf{y})$ denotes the volume occupied by the heat sink for a given parameter vector. 
The volume can be estimated using a simple analytical formula, given the geometrical characteristics of the heat sink.
Typically, there is no unique solution that minimizes both objectives at the same time. 
Instead, there exist several solutions which cannot be further improved with respect to one objective without simultaneously worsening the other.
These solutions are called Pareto optimal and form a so-called Pareto front. 
Once the Pareto front is available, the design engineer can choose the final heat sink designs according to his/her expertise.

Note that, as discussed in Section~\ref{subsec:qoi}, the steady-state temperature of the heat sink $T_{\text{s}}$ is directly connected to the maximum junction temperature $T_{\text{j}}$ via formulas \eqref{eq:thermal_resistance} and \eqref{eq:junction_temp}. 
Therefore, all aforementioned optimization problems can be alternatively formulated such that a minimal value for $T_{\text{s}}$ is sought instead of $T_{\text{j}}$.

There exist numerous methods and algorithms for solving the aforementioned optimization problems.
In the context of engineering design, stochastic optimization methods \cite{schneider2007stochastic} are particularly popular due to the fact that they employ computational models in a black-box manner and avoid convergence problems due to local optima, which are often encountered when using deterministic optimization methods.
On the downsides, stochastic optimization algorithms typically require a large number of model evaluations until converging to the sought optimal solutions, which can lead to severe computational costs, particularly if the employed model is time and resource demanding.
Hence the necessity for an inexpensive albeit sufficiently accurate surrogate model, which is the focus of Section~\ref{sec:ddpce}.

\section{Data-driven polynomial chaos expansion}
\label{sec:ddpce}

\subsection{Polynomial chaos expansion}
\label{subsec:pce}

We assume that the functional dependence between a scalar \gls{qoi} and an $N$-dimensional vector of features $\mathbf{y} = \left(y_1, \dots, y_N\right)$ is described by the deterministic map $S: \mathbb{R}^N \rightarrow \mathbb{R}$.
In the context of this work, the feature vector refers to the operational and geometrical heat sink design features such that $\mathbf{y} = \left(l, g_{\text{f}}, w_{\text{f}}, h_{\text{f}}, h_{\text{b}}, N_{\text{f}}, v, T_{\text{a}}, P\right)$, the map $S$ refers to evaluating and post-processing the \gls{rans}-based \gls{cfd} simulation model (see Section~\ref{subsec:simulation}), and the \gls{qoi} is the heat sink's temperature $T_{\text{s}}$, which can then be used to infer the junction temperature $T_{\text{j}}$ using formulas \eqref{eq:thermal_resistance} and \eqref{eq:junction_temp}. 

A \gls{pce} is a global polynomial approximation of the form
\begin{equation}
	\label{eq:pce}
	S(\mathbf{y}) \approx \widetilde{S}(\mathbf{y}) = \sum_{k=1}^K s_k \Psi_{k}\left(\mathbf{y}\right),
\end{equation}
where $s_k$ are scalar coefficients and $\Psi_k$ are multivariate orthonormal polynomials, such that 
\begin{equation}
	\label{eq:orthNd}
	\int_{\mathbb{R}^N} \Psi_{k}\left(\mathbf{y}\right) \Psi_{l}\left(\mathbf{y}\right) \varrho\left(\mathbf{y}\right) \mathrm{d}\mathbf{y} =  \delta_{kl},
\end{equation}
with $\delta_{kl}$ being the Kronecker delta and $\varrho\left(\mathbf{y}\right)$ the joint \gls{pdf} that collectively characterizes the design features \cite{sudret2008global}.
Assuming that feature realizations occur independently from one another, the joint \gls{pdf} is given as
\begin{align}
	\varrho\left(\mathbf{y}\right) &= \prod_{n=1}^N \varrho_n\left(y_n\right), \label{eq:joint_pdf}
\end{align}
where $\varrho_n\left(y_n\right)$, $n=1, \dots, N$, are univariate \glspl{pdf} corresponding to single features \cite{migliorati2014analysis}. 
We note that the independence assumption is not crucial, i.e. dependent parameters can be addressed with suitable transformations \cite{feinberg2018}.
Nevertheless, it has been shown that the independence assumption often leads to improved results in the context of data-driven approximation via \glspl{pce} \cite{torre2019data}.
For the purpose of this study, the design features are modeled as uniformly distributed within the corresponding bounds given in Table~\ref{tab:features}, i.e. they may take any value within these bounds with equal probability.
The multivariate polynomials are then given by
\begin{equation}
	\label{eq:multivariate-poly}
	\Psi_k\left(\mathbf{y}\right) = \Psi_{\mathbf{p}}\left(\mathbf{y}\right) = \prod_{n=1}^N \psi_{n}^{p_n}\left(y_n\right),
\end{equation}
where $\psi_n^{p_n}$ are univariate Legendre polynomials in $y_n$ with degree $p_n$ \cite{loukrezis2020robust}. The univariate Legendre polynomials are orthonormal with respect to the uniform univariate \glspl{pdf} $\varrho_n\left(y_n\right)$ \cite{xiu2002wiener}. 
The multi-index $\mathbf{p} = \left(p_1, \dots, p_N\right)$ collects the polynomial degrees of the univariate polynomials that comprise the multivariate polynomial $\Psi_{\mathbf{p}}$ in \eqref{eq:multivariate-poly}.
There exists a one-to-one relation between the multi-indices $\mathbf{p}$ and the global indices $k=1,\dots,K$ in \eqref{eq:pce}, such that
\begin{equation}
	\label{eq:pce_multiidx}
	\widetilde{S}(\mathbf{y}) = \sum_{k=1}^K s_k \Psi_{k}\left(\mathbf{y}\right) = \sum_{\mathbf{p} \in \Lambda} s_{\mathbf{p}} \Psi_{\mathbf{p}}\left(\mathbf{y}\right),
\end{equation}
where $\Lambda$ is a multi-index set with cardinality $\#\Lambda = K$. 

\subsection{Polynomial basis and model structure}
\label{subsec:structure}
There exist numerous options regarding the model structure of a \gls{pce}, equivalently, for choosing the multi-indices that comprise the multi-index set $\Lambda$ and thus form the polynomial basis $\left\{\Psi_{\mathbf{p}}\right\}_{\mathbf{p} \in \Lambda}$.
One obvious choice is to construct the multivariate polynomial basis as a tensor product of $N$ univariate polynomial bases, such that
\begin{equation}
	\left\{\Psi_{\mathbf{p}}\right\}_{\mathbf{p} \in \Lambda} = \left\{\psi_{1}^{p_1}\right\}_{p_1 = 1}^{P_1} \otimes \cdots \otimes \left\{\psi_{N}^{p_N}\right\}_{p_N = 1}^{P_N},
\end{equation}
where $\otimes$ denotes a tensor product and $P_n \in \mathbb{Z}_{\geq 0}$, $n=1, \dots, N$.
Assuming that $P_n=P$, $n=1,\dots,N$, the corresponding multi-index set is given by
\begin{equation}
	\label{eq:tp_set}
	\Lambda_\mathrm{TP}^P = \left\{\mathbf{p} : \max_{n=1,\dots,N}\left\{p_n\right\} \leq P \right\},
\end{equation}
and the cardinality of the basis is equal to $\#\Lambda_{\text{TP}}^P = P^N$.
Due to the exponential dependence of the basis size on the number of features, tensor-product bases become intractable even for relatively small values of $P$ and $N$. 

Motivated by the so-called ``sparsity of effects'' principle \cite{blatman2011adaptive}, i.e. the assumption that only few feature interactions have a non-negligible impact on the \gls{qoi}, additional constraints on the model structure can be placed.
A popular choice is to use a so-called total-degree \gls{pce} basis, which comprises polynomials corresponding to the multi-index set
\begin{equation}
	\label{eq:td_set}
	\Lambda_\mathrm{TD}^P = \left\{\mathbf{p} : \sum_{n=1}^N p_n \leq P \right\},
\end{equation}
where $P \in \mathbb{Z}_{\geq 0}$, in which case the cardinality of the \gls{pce} basis is reduced to $\#\Lambda_{\text{TD}}^P = \frac{\left(N+P\right)!}{N! P!}$.
The number of basis terms can be reduced even further if a hyperbolic basis \cite{blatman2011adaptive} is used, corresponding to the multi-index set
\begin{equation}
	\label{eq:hyp_set}
	\Lambda_\mathrm{H}^P = \left\{\mathbf{p} : \left(\sum_{n=1}^N \left(p_n\right)^q\right)^{1/q} \leq P\right\},
\end{equation}
where $0<q<1$.
Note that for $q=1$, the total-degree basis given in \eqref{eq:td_set} is retrieved. 

\glspl{pce} based on tensor-product, total-degree, or hyperbolic polynomial bases are models of fixed structure. 
Equivalently, the multi-index set $\Lambda$ is a priori chosen.
This can be a limiting factor due to the fact that a fixed model may contain terms that do not affect significantly the \gls{qoi} and could therefore be omitted. 
Accordingly, a fixed model may neglect influential terms which could improve its predictive capability if added.
Additionally, these bases are isotropic, i.e. they handle the input features in a similar manner, while in most practical cases the features have a different impact on the \gls{qoi} which can be taken into consideration to form an anisotropic basis.
Therefore, numerous approaches for constructing so-called sparse \glspl{pce} that neglect non-influential basis terms have been proposed, where the \gls{pce} coefficients are computed using either numerical quadrature \cite{ahlfeld2016samba, conrad2013adaptive, constantine2012sparse, winokur2016sparse} or regression \cite{abraham2017robust, blatman2010adaptive, blatman2011adaptive, cheng2018sparse, diaz2018sparse, doostan2011non, hampton2018basis, he2020adaptive, hu2011adaptive, huan2018compressive, loukrezis2020robust, migliorati2013a, peng2014weighted, tsilifis2019compressive, zhao2019efficient}.
The latter approach is essential for \glspl{ddpce}, as it recasts the \gls{pce} as a supervised \gls{ml} method based on polynomial regression, as discussed next.

\subsection{Polynomial chaos regression}
\label{subsec:coeff}
We now assume that a fixed polynomial basis $\left\{\Psi_{\mathbf{p}}\right\}_{\mathbf{p} \in \Lambda}$, $\#\Lambda = K$, and a dataset $\mathcal{D} = \left\{\left(\mathbf{y}^{(m)}, S\left(\mathbf{y}^{(m)}\right) \right) \right\}_{m=1}^M$ containing $M$ realizations of the feature vector and the corresponding \gls{qoi} are available. 
This dataset may consist of measurement or simulation-generated data, e.g. by evaluating the model presented in Section~\ref{subsec:simulation} for different design configurations. 
The \gls{pce} coefficients $\mathbf{s} = \left(s_{\mathbf{p}}\right)_{\mathbf{p} \in \Lambda}$ can then be computed by means of regression \cite{vapnik2013nature, ratner2017statistical}, as follows.

If $M>K$, the coefficient vector $\mathbf{s}$ is obtained by solving the discrete \gls{ls} minimization problem
\begin{equation}
	\label{eq:ls_min_problem}
	\mathbf{s} = \argmin_{\hat{\mathbf{s}} \in \mathbb{R}^{K}} \left\{ \sum_{m=1}^M \left(S\left(\mathbf{y}^{(m)}\right) - \sum_{\mathbf{p} \in \Lambda} \hat{s}_{\mathbf{p}} \Psi_{\mathbf{p}}\left(\mathbf{y}^{(m)}\right) \right)^2\right\}.
\end{equation}
To solve \eqref{eq:ls_min_problem}, we introduce the so-called design matrix $\mathbf{D} \in \mathbb{R}^{M \times K}$ with elements $d_{mk}=\Psi_{k}\left(\mathbf{y}^{(m)}\right)$ and the \gls{qoi} value vector  $\mathbf{b}=\left(S\left(\mathbf{y}^{(1)}\right), \dots, S\left(\mathbf{y}^{(M)}\right)\right)$. 
The discrete minimization problem \eqref{eq:ls_min_problem} can then be written in the equivalent algebraic form
\begin{equation}
	\label{eq:min_algebraic}
	\mathbf{s} = \argmin_{\hat{\mathbf{s}} \in \mathbb{R}^K}\left\{ \|\mathbf{D} \hat{\mathbf{s}} - \mathbf{b} \|_2^2 \right\}.
\end{equation}
Applying the necessary conditions for a minimum, we obtain the solution
\begin{equation}
	\label{eq:normal_solution}
	\mathbf{s}  = \left(\mathbf{D}^\top \mathbf{D}\right)^{-1} \mathbf{D}^\top \mathbf{b}.
\end{equation}
To avoid conditioning problems, the solution is more commonly obtained with a QR decomposition of the design matrix $\mathbf{D}$ \cite{higham2002}.
The conditioning of the \gls{ls} problem can be further improved with the use of optimal experimental design methods \cite{loukrezis2020robust, loukrezisPhD, diaz2018sparse, hadigol2018least, hampton2015coherence, fajraoui2017sequential, zein2013efficient}.

If $M<K$, the \gls{ls} problem \eqref{eq:ls_min_problem} is ill-posed.
Moreover, even if $M>K$, but with $M \approx K$, the \gls{ls} problem will most probably suffer from conditioning issues leading to an overfitted solution \cite{migliorati2014analysis}.  
In that case, a regularized \gls{ls} problem is solved instead, given as 
\begin{align}
	\label{eq:ls-reg}
	\mathbf{s} = \argmin_{\hat{\mathbf{s}} \in \mathbb{R}^{K}} \left\{ \sum_{m=1}^M \left(S\left(\mathbf{y}^{(m)}\right) - \sum_{\mathbf{p} \in \Lambda} \hat{s}_{\mathbf{p}} \Psi_{\mathbf{p}}\left(\mathbf{y}^{(m)}\right) \right)^2 + \lambda R\left(\hat{\mathbf{s}}\right)\right\},
\end{align}
where the loss function of \eqref{eq:ls_min_problem} is modified with the addition of a regularization function $R\left(\cdot\right)$, which imposes some form of constraint upon the coefficient vector, and a weighting parameter $\lambda$ which controls the importance of the regularization in the minimization.
Specifically for computing \glspl{pce}, it is common to use $\ell_1$-regularization, such that $R\left(\hat{\mathbf{s}}\right) = \left\|\hat{\mathbf{s}}\right\|_1$ \cite{doostan2011non,  peng2014weighted, tsilifis2019compressive, hampton2015compressive}, a technique closely connected to the \gls{lasso} method in statistics \cite{tibshirani1996regression} and to compressed sensing methods in signal processing \cite{donoho2006compressed}. 
The $\ell_1$-regularization forces multiple coefficients to zero, thus effectively reducing the model parameters and resulting in a sparse \gls{pce}.
The $\ell_1$-regularized \gls{ls} problem is commonly solved with linear programming algorithms \cite{becker2011nesta, candes2006stable, van2009probing} or with the \gls{lar} algorithm \cite{blatman2011adaptive, efron2004least}. 
The latter approach can additionally be complemented with optimal experimental design and active learning methods \cite{fajraoui2017sequential, marelli2018active}.

\section{Model-form uncertainty, model robustness, and ensemble modeling}
\label{sec:ensemble}
As previously noted, the \gls{ddpce} is a supervised \gls{ml} method.
As is common in the context of supervised \gls{ml}, to train and assess a data-driven model, an initial dataset of feature realizations and corresponding \gls{qoi} observations is separated into a training dataset $\mathcal{D}_{\text{train}} = \left\{\left(\mathbf{y}^{(m)}, S\left(\mathbf{y}^{(m)}\right)\right) \right\}_{m=1}^M$ and a test dataset $\mathcal{D}_{\text{test}} = \left\{\left(\mathbf{y}^{(j)}, S\left(\mathbf{y}^{(j)}\right)\right) \right\}_{j=1}^J$, such that $\mathcal{D}_{\text{train}} \cap \mathcal{D}_{\text{test}} = \emptyset$.
The training dataset is here employed to compute the \gls{pce} coefficients by means of regression, i.e. for solving the problem \eqref{eq:ls_min_problem}, respectively \eqref{eq:ls-reg}. 
The test dataset is then used to evaluate the prediction accuracy of the trained model on previously unseen data, e.g. using a suitable error norm.

However, the performance of data-driven surrogate models is crucially dependent on which subsets of the full dataset are selected as training or test data.  
That is, a specific partition of the dataset to training and test data might lead to results that are not representative of the actual predictive capability of the model.
Exemplarily, in the case of \glspl{ddpce}, different training datasets result in variations in the \gls{pce} coefficients, thus affecting the model's predictions. 
For sparse \glspl{pce} in particular, different training datasets can lead to variations in the number of non-zero coefficients, subsequently to the number of expansion terms.
This results in a so-called epistemic or model-form uncertainty \cite{der2009aleatory}, which must be accounted for when providing surrogate-based predictions.

To avoid this pitfall, in this work we resort to the solution of ensemble modeling \cite{goel2007ensemble, ren2022ensemble}. 
That is, the original dataset is randomly re-ordered $I$ times, thus generating training and test datasets $\mathcal{D}_{\text{train}}^{(i)}$, $\mathcal{D}_{\text{test}}^{(i)}$, $i=1,\dots,I$, accordingly, an ensemble of surrogate models $\left\{\widetilde{S}^{(i)}\left(\mathbf{y}\right)\right\}_{i=1}^I$. 
Aggregating the results over the $I$ dataset combinations and corresponding surrogate models, one can then compute mean values and standard deviations for the error metrics which are used to evaluate the surrogate model's accuracy.
In that way, we are not only able to assess the performance of a surrogate model with respect to a specific accuracy metric on average, but also evaluate its robustness against variations in the training and test data.
Once available, all surrogate models $\widetilde{S}(\mathbf{y})$, $i=1,\dots,I$, can be employed for predicting the \gls{qoi} for a given feature combination, such that the \gls{qoi} prediction takes the form of a mean value along with a standard deviation or a confidence interval, thus quantifying the prediction uncertainty attributed to the uncertainty in the model.
It is expected that model uncertainty and the corresponding uncertainty in the data-driven model's predictions will be reduced as the training dataset is increased in size, as also verified by the numerical results presented in Section~\ref{sec:applic-num-res}.

\section{Numerical Results}
\label{sec:applic-num-res}

\subsection{Predictive capability of data-driven polynomial chaos expansions}
\label{subsec:predictive}

The numerical experiments presented in this section aim to assess the predictive capabilities of \glspl{ddpce}, as well as compare them against other data-driven surrogate models, specifically \glspl{gp} \cite{williams2006gaussian} and \glspl{nn} \cite{graupe2013principles}. 
Surrogate models based on \glspl{svm} were also tested, but their predictive capability was found to be significantly worse than the other surrogate models, therefore, \gls{svm}-based results have been omitted.
The compared surrogate models are described next.
\begin{itemize}
	\item Total-degree \glspl{ddpce} for a maximum polynomial order of $P=3$ are constructed according to formula \eqref{eq:td_set}. The choice of $P=3$ yielded superior results compared to total-degree bases with different maximum polynomial orders. In this case, the polynomial basis always consists of $220$ terms.
	\item Sparse \glspl{ddpce} are obtained with the sensitivity-adaptive algorithm developed by the authors in \cite{loukrezis2020robust}. 
		In particular, the polynomial basis is adaptively expanded by including polynomial terms corresponding to large \gls{pce} coefficients, which in turn are connected to variance-based sensitivity indices \cite{crestaux2009polynomial, sudret2008global}. The basis expansion procedure continues as long as the condition number of the design matrix $\mathbf{D}$ remains smaller than $10$. This stopping criterion was found to perform better than other options. 
		In this case, the size of the polynomial basis varies depending on the training dataset, as will be discussed next.
	\item The \gls{gp} surrogate model employs a Mat\'ern covariance function, also referred to as the ``kernel'', with a Mat\'ern coefficient $\nu=5/2$. This choice was found to outperform other popular kernel options, e.g. (squared) exponential kernels or Mat\'ern kernels with different coefficient values.
	\item The \gls{nn} surrogate model consists of $2$ hidden layers with $10$ neurons per layer. Hyperbolic tangent activation functions are employed and a limited memory BFGS optimizer is used for minimizing the loss function. These options with respect to the architecture of the \gls{nn} and its training were found to outperform several other tested alternatives.
\end{itemize}

To train and assess the accuracy of the surrogate models, we use a dataset with $935$ readily available feature value combinations and the corresponding heat sink temperatures $T_{\text{s}}$. 
These results have been obtained using the ANSYS Icepak simulation model presented in Section~\ref{subsec:simulation} for random realizations of the design features within the value ranges given in Table~\ref{tab:features}.
Note that the average duration of a single simulation on an up-to-date machine is $35$ minutes.
We compute surrogate models using training datasets of increasing size, i.e. $M=100, 200, \dots, 800$. 
The test dataset has a fixed size of $J=135$ data points, which are not included in the training data.
For the ensemble modeling procedure described in Section~\ref{sec:ensemble}, the original dataset is randomly re-ordered $I=100$ times.

The performance of a surrogate model in terms of predictive capability is assessed using the following error metrics.
Denoting the output of the original model with $S(\mathbf{y})$ and that of a surrogate model with $\widetilde{S}\left(\mathbf{y}\right)$, the relative prediction error on the $j$-th test data point is 
\begin{equation}
	\epsilon_{\text{rel}, j} = \left| \frac{S\left(\mathbf{y}^{(j)}\right) - \widetilde{S}\left(\mathbf{y}^{(j)}\right)}{S\left(\mathbf{y}^{(j)}\right)} \right|.
\end{equation}
The average performance of the surrogate model in terms of prediction accuracy is quantified using the mean absolute percentage error
\begin{equation}
	\label{eq:mean_rel_err}
	\epsilon_{\text{mean}} = \frac{1}{J}\sum_{j=1}^J \epsilon_{\text{rel},j}.
\end{equation}
To obtain a worst-case performance estimate, we compute the maximum absolute percentage error
\begin{equation}
	\label{eq:max_rel_err}
	\epsilon_{ \max} = \max_{j=1,\dots,J} \left\{\epsilon_{\text{rel}, j}\right\}.
\end{equation}
Aggregating the error results over the $I=100$ dataset combinations and corresponding surrogate models, we compute the mean values $\mu\left(\cdot\right)$ and the standard deviations $\sigma\left(\cdot\right)$ of the errors \eqref{eq:mean_rel_err} and \eqref{eq:max_rel_err}.
Moreover, an absolute worst-case prediction estimate is computed as $\max_i\left\{\epsilon_{\max}^{(i)}\right\}$, $i=1,\dots,100$.

\begin{figure}[t!]
	\begin{subfigure}[t]{0.33\textwidth}
		\begin{tikzpicture}
			\begin{semilogyaxis}[xlabel={Training dataset size $M$}, ylabel={$\mu\left(\epsilon_{\text{mean}}\right)$}, width=1\textwidth, y label style={yshift=-0.5em}, x label style={yshift=0.5em}, label style={font=\normalsize}, tick label style={font=\normalsize}, legend pos=north east, legend style={font=\normalsize}, ymax=1.0, ymin=0.001]
				\addplot[mark=o, color=black] table[x index=0, y index=1]{err_stats/err_stats_apce_Tj.txt};
				\addplot[mark=*, color=black, dashed, mark options={solid}] table[x index=0, y index=1]{err_stats/err_stats_tdpce_Tj.txt};
				\addplot[mark=o, color=gray] table[x index=0, y index=1]{err_stats/err_stats_gpr_Tj.txt};
				\addplot[mark=*, color=gray, dashed, mark options={solid}] table[x index=0, y index=1]{err_stats/err_stats_nn_Tj.txt};
			\end{semilogyaxis}
		\end{tikzpicture}
	\end{subfigure}
	\begin{subfigure}[t]{0.33\textwidth}
		\begin{tikzpicture}
			\begin{semilogyaxis}[xlabel={Training dataset size $M$}, ylabel={$\mu\left(\epsilon_{\max}\right)$}, width=1\textwidth, y label style={yshift=-0.5em}, x label style={yshift=0.5em}, label style={font=\normalsize}, tick label style={font=\normalsize}, legend pos=north east, legend style={font=\normalsize}, ymax=10, ymin=0.01]
				\addplot[mark=o, color=black] table[x index=0, y index=3]{err_stats/err_stats_apce_Tj.txt};
				\addplot[mark=*, color=black, dashed, mark options={solid}] table[x index=0, y index=3]{err_stats/err_stats_tdpce_Tj.txt};
				\addplot[mark=o, color=gray] table[x index=0, y index=3]{err_stats/err_stats_gpr_Tj.txt};
				\addplot[mark=*, color=gray, dashed, mark options={solid}] table[x index=0, y index=3]{err_stats/err_stats_nn_Tj.txt};
			\end{semilogyaxis}
		\end{tikzpicture}
	\end{subfigure}
	\begin{subfigure}[t]{0.33\textwidth}
		\begin{tikzpicture}
			\begin{semilogyaxis}[xlabel={Training dataset size $M$}, ylabel={$\max_i\left\{\epsilon_{\max}^{(i)}\right\}$}, width=1\textwidth, y label style={yshift=-0.5em}, x label style={yshift=0.5em}, label style={font=\normalsize}, tick label style={font=\normalsize}, legend pos=north east, legend style={font=\normalsize}, ymax=10, ymin=0.01]
				\addplot[mark=o, color=black] table[x index=0, y index=5]{err_stats/err_stats_apce_Tj.txt};
				\addplot[mark=*, color=black, dashed, mark options={solid}] table[x index=0, y index=5]{err_stats/err_stats_tdpce_Tj.txt};
				\addplot[mark=o, color=gray] table[x index=0, y index=5]{err_stats/err_stats_gpr_Tj.txt};
				\addplot[mark=*, color=gray, dashed, mark options={solid}] table[x index=0, y index=5]{err_stats/err_stats_nn_Tj.txt};
			\end{semilogyaxis}
		\end{tikzpicture}
	\end{subfigure}
	\\
	\begin{subfigure}[t]{0.33\textwidth}
		\begin{tikzpicture}
			\begin{semilogyaxis}[xlabel={Training dataset size $M$}, ylabel={$\sigma\left(\epsilon_{\text{mean}}\right)$}, width=1\textwidth, y label style={yshift=-0.5em}, x label style={yshift=0.5em}, label style={font=\normalsize}, tick label style={font=\normalsize}, legend pos=north east, legend style={font=\normalsize}, ymax=0.1, ymin=0.0001]
				\addplot[mark=o, color=black] table[x index=0, y index=2]{err_stats/err_stats_apce_Tj.txt};
				\addplot[mark=*, color=black, dashed, mark options={solid}] table[x index=0, y index=2]{err_stats/err_stats_tdpce_Tj.txt};
				\addplot[mark=o, color=gray] table[x index=0, y index=2]{err_stats/err_stats_gpr_Tj.txt};
				\addplot[mark=*, color=gray, dashed, mark options={solid}] table[x index=0, y index=2]{err_stats/err_stats_nn_Tj.txt};
			\end{semilogyaxis}
		\end{tikzpicture}
	\end{subfigure}
	\begin{subfigure}[t]{0.33\textwidth}
		\begin{tikzpicture}
			\begin{semilogyaxis}[xlabel={Training dataset size $M$}, ylabel={$\sigma\left(\epsilon_{\max}\right)$}, width=1\textwidth, y label style={yshift=-0.5em}, x label style={yshift=0.5em}, label style={font=\normalsize}, tick label style={font=\normalsize},  legend style={font=\normalsize, at={(1.2,0.5)}, anchor=west}, ymax=1, ymin=0.001]
				\addplot[mark=o, color=black] table[x index=0, y index=4]{err_stats/err_stats_apce_Tj.txt};
				\addplot[mark=*, color=black, dashed, mark options={solid}] table[x index=0, y index=4]{err_stats/err_stats_tdpce_Tj.txt};
				\addplot[mark=o, color=gray] table[x index=0, y index=4]{err_stats/err_stats_gpr_Tj.txt};
				\addplot[mark=*, color=gray, dashed, mark options={solid}] table[x index=0, y index=4]{err_stats/err_stats_nn_Tj.txt};
				\legend{\small{DD-PCE (sparse-adaptive)}, \small{DD-PCE (total-degree)}, \small{GP}, \small{NN}}
			\end{semilogyaxis}
		\end{tikzpicture}
	\end{subfigure}
	\caption{Performance metrics for each surrogate model in relation to the size $M$ of the training dataset. The test dataset has a fixed size of $J = 135$ data points.}
	\label{fig:errors-vs-dataset-size_Ts}
\end{figure}
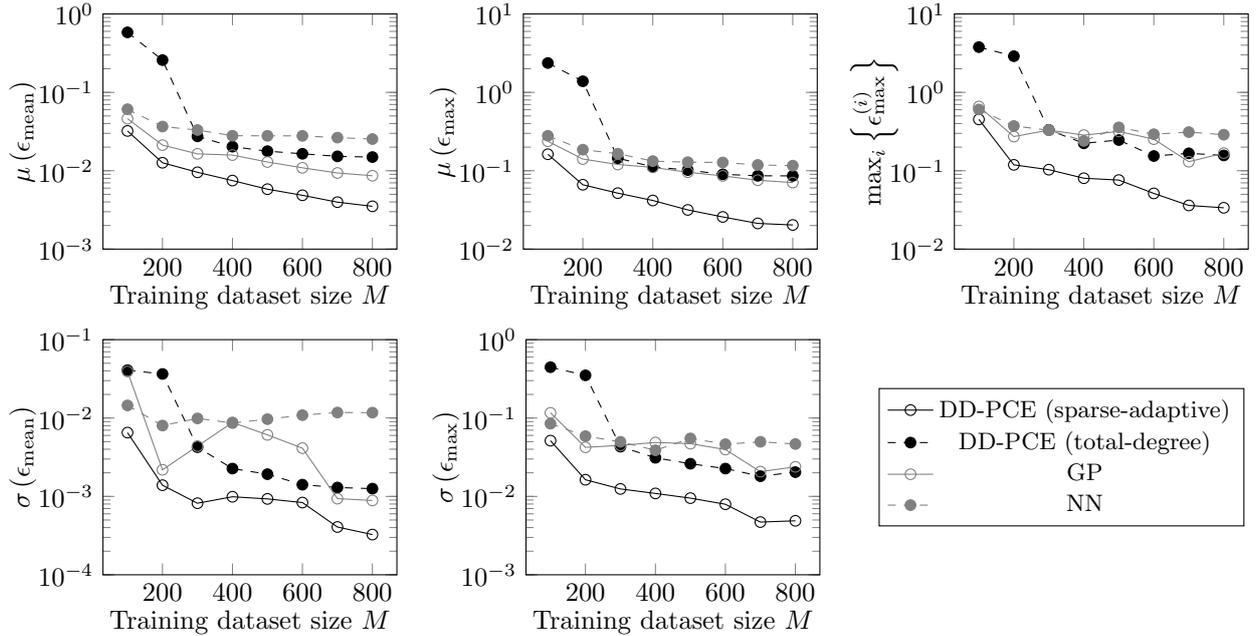

The corresponding numerical results are presented in Figure~\ref{fig:errors-vs-dataset-size_Ts}.
It can be observed that sparse-adaptive \glspl{ddpce} outperform all other surrogate models for all training dataset sizes, both in terms of accuracy and robustness, as they yield a superior predictive accuracy both on average and in the worst-case, while the corresponding error standard deviation values are significantly smaller.
Comparing the errors of sparse-adaptive and total-degree \glspl{ddpce}, it is clear that the flexibility in terms of model structure offered by the former plays a crucial role in both prediction accuracy and robustness.
Of particular interest is the observation that sparse-adaptive \glspl{ddpce} trained with as few as $M=200$ data points reach error values that cannot be reached by the other surrogate models even if trained with datasets four times larger.
We can therefore conclude that, in the considered setting, sparse-adaptive \glspl{ddpce} are particularly suitable for computing accurate surrogate models, even if the available training datasets are relatively small.

Focusing on sparse-adaptive \glspl{ddpce}, Table~\ref{tab:sparse-adaptive-pce-stats} presents the mean basis sizes and their standard deviations, along with the minimum and maximum basis sizes, which are computed out of the $100$ sparse-adaptive \glspl{ddpce} constructed for each size of the training dataset. As mentioned before, the size of the polynomial basis is not fixed in this case, but instead depends on the employed training dataset. This is due to the fact that the training data are employed in the construction of the design matrix $\mathbf{D}$, thus affecting its condition number. The latter is used as stopping criterion for the adaptive basis expansion. As previously noted in Section~\ref{sec:ensemble}, the different training datasets have a major impact on the data-driven model, resulting in significant differences in terms of model structure. These results consolidate why it is necessary to assess the robustness of the data-driven model and use the full ensemble for surrogate-based predictions accompanied by suitable uncertainty metrics.

\begin{table}[b!]
	\centering
	\caption{Mean values, standard deviations, minimum values, and maximum values for the size of the polynomial bases of sparse-adaptive \glspl{ddpce} computed for each size of the training dataset.}
	\begin{tabular}{|l|c|c|c|c|c|c|c|c|}
		\hline 
		Training dataset size $M$ & 100 & 200 & 300 & 400 & 500 & 600 & 700 & 800 \\ 
		\hline
		Basis size, mean & 63.8 & 110.3 & 152.1 & 194.4 & 234.0 & 268.3 & 300.9 & 329.3\\
		\hline
		Basis size, st. dev. & 3.8 & 6.1 & 9.4 & 10.6 & 13.5 & 13.9 & 14.3 & 14.7\\
		\hline
		Basis size, min. & 51 & 95 & 123 & 172 & 189 & 218 & 262 & 291\\
		\hline 
		Basis size, max. & 71 & 123 & 176 & 223 & 257 & 293 & 331 & 369\\
		\hline
	\end{tabular}
	\label{tab:sparse-adaptive-pce-stats}
\end{table}

\subsection{Surrogate-based stochastic optimization}
\label{subsec:surrogate-based-optimization}
In the following, we assume that the considered power module operates at a fixed ambient temperature $T_{\text{a}} = 45$~\si{\degreeCelsius} and with constant power losses $P = 140$~\si{\watt}, i.e. the maximum allowed values for these two features (see Table~\ref{tab:features}). 
In the first optimization study, shown in Section~\ref{subsubsec:single-obj-optimization}, we aim to identify heat sink designs that minimize the heat sink temperature $T_{\text{s}}$, thus the junction temperature $T_\text{j}$ as well, see Sections~\ref{subsec:qoi} and \ref{subsec:optimization}.
To that end, the single-objective optimization problem \eqref{eq:single-objective-opt} is solved using the \gls{pso} \cite{kennedy1995particle, poli2007particle} and \gls{de} \cite{price2006differential} algorithms.
In the second optimization study, shown in Section~\ref{subsubsec:multi-obj-optimization}, two design objectives are considered as in the multi-objective optimization problem \eqref{eq:multi-objective-opt}, such that the junction temperature $T_{\text{j}}$ and the heat sink's volume $V_{\text{s}}$ are simultaneously minimized. 
In this case we use a \gls{nsga} algorithm \cite{deb2002fast, deb2006reference}.
All optimization algorithms are implemented in the Python package \texttt{pymoo} \cite{blank2020pymoo}.
For both optimization problems, the sparse-adaptive \gls{ddpce} surrogate models are employed within the optimization algorithms, as they were found to outperform the competitive surrogate models in Section~\ref{subsec:predictive}.

\subsubsection{Single-objective surrogate-based optimization}
\label{subsubsec:single-obj-optimization}
By using the sparse-adaptive \gls{ddpce} surrogate models $\left\{\widetilde{S}^{(i)}_{M}(\mathbf{y})\right\}_{i=1}^{I=100}$, trained with datasets of increasing size $M=100, 200, \dots, 800$, (see Section~\ref{subsec:predictive}), the optimization problem \eqref{eq:single-objective-opt} is effectively solved $I$ times per training dataset size, i.e. once with each corresponding surrogate. 
Then, by aggregating the $I$ optimization results, the minimum heat sink temperature $T_{\text{s}, \min}$ is given in the form of a mean value and a corresponding standard deviation.
The same holds for the optimal design parameter values, denoted in the following as $l^*$, $g_{\text{f}}^*$, $w_{\text{f}}^*$, $h_{\text{f}}^*$, $h_{\text{b}}^*$, $N_{\text{f}}^*$, and $v^*$.
Denoting the mean and standard deviation values of $T_{\text{s}, \min}$ with $\mu\left(T_{\text{s}, \min}\right)$ and $\sigma\left(T_{\text{s}, \min}\right)$, respectively, a worst-case estimate for the minimum junction temperature $T_{\text{j}, \min}$ is obtained by setting $T_{\text{s}, \min} = \mu\left(T_{\text{s}, \min}\right) + 6 \sigma\left(T_{\text{s}, \min}\right)$, as is often done in the context of reliability engineering to ensure that the design will comply to the requirements with a very high probability. 
Such designs are commonly referred to as six-sigma $(6\sigma)$ \cite{koch2004design}.

\begin{table}[b!]
	\centering
	\caption{Single-objective constrained optimization with the \gls{pso} algorithm: minimum temperature values and computational cost.}
	\begin{tabular}{|l|c|c|c|c|c|c|c|c|}
		\hline 
		Training dataset size $M$ & 100 & 200 & 300 & 400 & 500 & 600 & 700 & 800 \\ 
		\hline
		Average model evaluations & 1639 & 1696 & 1681 & 1598 & 1542 & 1600 & 1603 & 1734\\
		\hline
		$T_{\text{s}, \min}$ (\si{\degreeCelsius}), mean & 51.9 & 55.5 & 58.1 & 58.6 & 59.1 & 59.7 & 60.1 & 60.5 \\
		\hline
		$T_{\text{s}, \min}$ (\si{\degreeCelsius}), st. dev. & 5.4 & 2.6 & 1.7 & 1.7 & 1.2 & 1.2 & 0.9 & 0.7 \\
		\hline
		$T_{\text{j}, \min}$ (\si{\degreeCelsius}), six-sigma  & 187.9 & 174.7 & 171.9 & 172.4 & 169.9 & 170.5 & 169.1 & 168.3 \\
		\hline 
	\end{tabular}
	\label{tab:single-opt-pso-results}
\end{table}

\begin{table}[b!]
	\centering
	\caption{Single-objective constrained optimization with the \gls{de} algorithm: minimum temperature values and computational cost.}
	\begin{tabular}{|l|c|c|c|c|c|c|c|c|}
		\hline 
		Training dataset size $M$ & 100 & 200 & 300 & 400 & 500 & 600 & 700 & 800 \\ 
		\hline
		Average model evaluations & 6734 & 7672 & 7849 & 7505 & 7452 & 7299 & 7275 & 7090\\
		\hline
		$T_{\text{s}, \min}$ (\si{\degreeCelsius}), mean & 51.8 & 55.3 & 58.1 & 58.7 & 59.1 & 59.7 & 60.2 & 60.5 \\
		\hline
		$T_{\text{s}, \min}$ (\si{\degreeCelsius}), st. dev. & 5.4 & 2.6 & 1.7 & 1.6 & 1.2 & 1.1 & 0.8 & 0.7 \\
		\hline
		$T_{\text{j}, \min}$ (\si{\degreeCelsius}), six-sigma  & 187.8 & 174.8 & 171.9 & 171.9 & 169.9 & 169.9 & 168.6 & 168.3  \\
		\hline 
	\end{tabular}
	\label{tab:single-opt-DE-results}
\end{table}

Tables~\ref{tab:single-opt-pso-results} and \ref{tab:single-opt-DE-results} present the optimization results and the performance of the \gls{pso} and \gls{de} algorithms using \gls{ddpce} surrogate models of increasing accuracy, equivalently, \glspl{ddpce} trained with datasets of increasing size. 
A first observation is that the two optimization algorithms yield very similar results, however, the computational cost of the \gls{pso} algorithm, as measured by the number of model evaluations until an optimal design is identified, is significantly lower than the one of the \gls{de} algorithm. 
Nevertheless, even in the case of the \gls{pso} algorithm, the computational gains due to using the surrogate model instead of the original $3$D \gls{cfd} heat sink model are tremendous. 
Exemplarily, the (on average) $1734$ model evaluations with each of the $100$ \gls{ddpce} surrogate models computed with a training data set of size $M=800$ (see Table~\ref{tab:single-opt-pso-results}, last column), amount to $4.8$ CPU-hours of computation.
This computation time is exceeded by merely $9$ evaluations of the original heat sink model.

Considering the optimization results, it can be observed that using surrogate models of reduced accuracy results in an underestimation of the minimum heat sink temperature. 
Additionally, there is a much greater uncertainty attached to these results, as indicated by the corresponding standard deviation values.
There is a certain tendency to be observed, that is, as the size of the training dataset increases, along with the accuracy of the surrogate model, the estimations for $T_{\text{s}, \min}$ become more pessimistic. 
At the same time, the optimization results become increasingly more robust, as indicated by the progressively reduced standard deviation values. 
In the particular case of the least accurate surrogate model, i.e. for $M=100$, the uncertainty regarding the value of $T_{\text{s}, \min}$ leads to a six-sigma value for $T_{\text{j}, \min}$ which significantly exceeds the threshold of $175$~\si{\degreeCelsius} (also see Section~\ref{subsec:qoi}).
This constraint is respected for the remaining surrogate models and the margin to the threshold becomes more pronounced as the surrogate models increase in accuracy.
The corresponding optimal design parameter values, again given in the form of mean value and standard deviation, are shown in 
Tables~\ref{tab:single-opt-pso-results-params} and \ref{tab:single-opt-de-results-params} for the \gls{pso} and \gls{de} algorithms, respectively.
Once again, the \gls{pso} and \gls{de} algorithms yield very similar results, particularly for \glspl{ddpce} of increased accuracy.

\begin{table}[b!]
	\centering
	\caption{Single-objective constrained optimization with the \gls{pso} algorithm: optimal design parameters.}
	\begin{tabular}{|l|c|c|c|c|c|c|c|c|}
		\hline 
		Training dataset size $M$ & 100 & 200 & 300 & 400 & 500 & 600 & 700 & 800 \\ 
		\hline
		$h_{\text{b}}^*$ (\si{\milli\meter}), mean  & 14.2 & 14.8  & 15.0 & 14.9  & 15.0 & 15.0 & 14.8 & 14.8 \\
		$h_{\text{b}}^*$ (\si{\milli\meter}), st. dev. & 1.4 & 1.0 & 0.2 & 0.2 & 0.2 & 0.3 & 1.0 & 0.8 \\
		\hline
		$N_{\text{f}}^*$, mean  & 24.0 & 22.6 & 23.4 & 24.2 & 24.6 & 24.2 & 24.4 & 23.7 \\
		$N_{\text{f}}^*$, st. dev. & 2.8 & 2.5 & 2.4 & 1.8 & 1.0 & 1.1 & 0.8 & 1.1 \\
		\hline
		$h_{\text{f}}^*$ (\si{\milli\meter}), mean & 33.6 & 43.6 & 40.5 & 39.1 & 41.3 & 41.8 & 42.1 & 42.5 \\
		$h_{\text{f}}^*$ (\si{\milli\meter}), st. dev. & 8.8 & 6.1 & 7.9 & 7.9 & 4.7 & 4.1 & 3.0 & 2.8\\
		\hline
		$l^*$ (\si{\milli\meter}), mean & 183.2 & 185.9 & 170.1 & 184.2 & 185.5 & 189.4 & 189.5 & 192.8 \\
		$l^*$ (\si{\milli\meter}), st. dev. & 31.4 & 23.6 & 25.2 & 22.5 & 22.9 & 19.4 & 19.3 & 15.5 \\
		\hline
		$v^*$ (\si{\meter\per\second}), mean & 4.0 & 4.8 & 4.7 & 4.7 & 4.7 & 4.6 & 4.7 & 4.5\\
		$v^*$ (\si{\meter\per\second}), st. dev. & 1.2 & 0.6 & 0.5 & 0.5 & 0.5 & 0.6 & 0.6 & 0.6 \\
		\hline
		$g_{\text{f}}^*$ (\si{\milli\meter}), mean & 5.1 & 3.2 & 3.1 & 3.1 & 3.1 & 3.2 & 3.0 & 3.0 \\
		$g_{\text{f}}^*$ (\si{\milli\meter}), st. dev. & 2.3 & 1.1 & 0.7 & 0.5 & 0.5 & 0.8 & 0.05 & 0.05\\
		\hline
		$w_{\text{f}}^*$ (\si{\milli\meter}), mean & 3.1 & 3.2 & 3.9 & 4.0 & 3.9 & 4.0 & 4.0 & 3.9 \\
		$w_{\text{f}}^*$ (\si{\milli\meter}), st. dev. & 1.0 & 0.7 & 0.3 & 0.2 & 0.3 & 0.2 & 0.3 & 0.4 \\
		\hline
	\end{tabular}
	\label{tab:single-opt-pso-results-params}
\end{table}

\begin{table}[b!]
	\centering
	\caption{Single-objective constrained optimization with the \gls{de} algorithm: optimal design parameters.}
	\begin{tabular}{|l|c|c|c|c|c|c|c|c|}
		\hline 
		Training dataset size $M$ & 100 & 200 & 300 & 400 & 500 & 600 & 700 & 800 \\ 
		\hline
		$h_{\text{b}}^*$ (\si{\milli\meter}), mean  & 13.8 & 14.7  & 15.0 & 14.9  & 14.9 & 14.9 & 14.9 & 14.9 \\
		$h_{\text{b}}^*$ (\si{\milli\meter}), st. dev. & 1.6 & 0.9 & 0.2 & 0.3 & 0.1 & 0.5 & 0.4 & 0.3 \\
		\hline
		$N_{\text{f}}^*$, mean  & 24.2 & 21.9 & 22.7 & 23.9 & 24.6 & 24.2 & 24.3 & 23.7 \\
		$N_{\text{f}}^*$, st. dev. & 2.2 & 2.5 & 2.6 & 2.3 & 0.9 & 1.0 & 0.8 & 1.0 \\
		\hline
		$h_{\text{f}}^*$ (\si{\milli\meter}), mean & 33.1 & 44.3 & 42.2 & 40.4 & 41.0 & 42.0 & 41.9 & 42.0 \\
		$h_{\text{f}}^*$ (\si{\milli\meter}), st. dev. & 8.9 & 4.0 & 6.1 & 6.4 & 4.8 & 3.0 & 3.0 & 2.6\\
		\hline
		$l^*$ (\si{\milli\meter}), mean & 183.2 & 185.9 & 170.1 & 184.2 & 185.5 & 189.4 & 189.5 & 192.8 \\
		$l^*$ (\si{\milli\meter}), st. dev. & 31.1 & 13.8 & 21.4 & 21.9 & 17.5 & 17.6 & 13.6 & 12.3 \\
		\hline
		$v^*$ (\si{\meter\per\second}), mean & 4.2 & 4.9 & 4.7 & 4.7 & 4.7 & 4.7 & 4.7 & 4.7\\
		$v^*$ (\si{\meter\per\second}), st. dev. & 1.0 & 0.2 & 0.3 & 0.4 & 0.5 & 0.5 & 0.5 & 0.4 \\
		\hline
		$g_{\text{f}}^*$ (\si{\milli\meter}), mean & 4.5 & 3.6 & 3.1 & 3.0 & 3.1 & 3.1 & 3.0 & 3.0 \\
		$g_{\text{f}}^*$ (\si{\milli\meter}), st. dev. & 2.1 & 1.5 & 0.6 & 0.2 & 0.7 & 0.6 & 0.02 & 0.03\\
		\hline
		$w_{\text{f}}^*$ (\si{\milli\meter}), mean & 3.1 & 3.0 & 3.9 & 4.0 & 3.9 & 3.9 & 3.9 & 3.8 \\
		$w_{\text{f}}^*$ (\si{\milli\meter}), st. dev. & 1.0 & 0.8 & 0.4 & 0.2 & 0.3 & 0.2 & 0.3 & 0.4 \\
		\hline
	\end{tabular}
	\label{tab:single-opt-de-results-params}
\end{table}

\subsubsection{Multi-objective surrogate-based optimization}
\label{subsubsec:multi-obj-optimization}

\begin{figure}[t!]
	\includegraphics[width=0.45\textwidth]{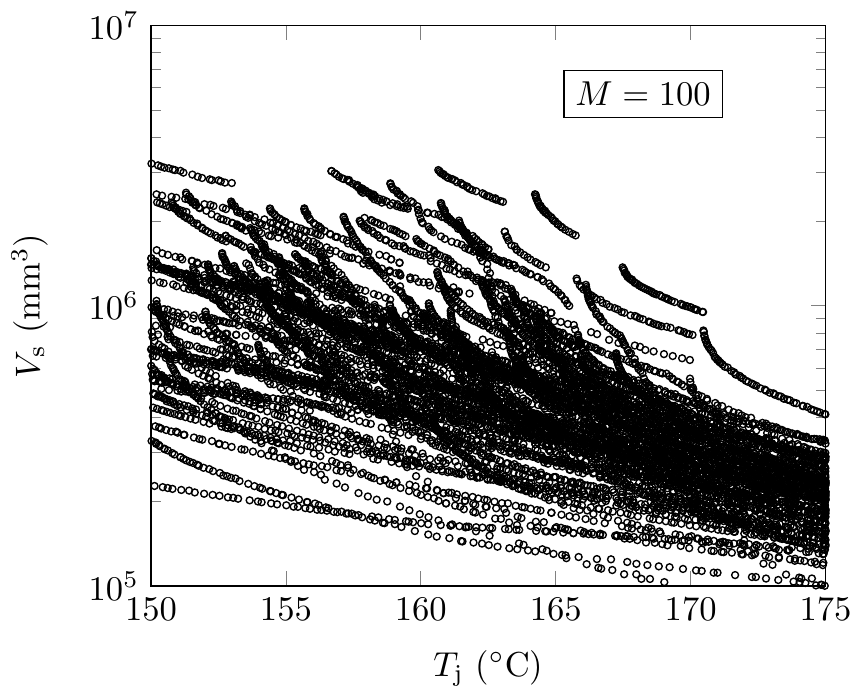}
	\hfill
	\includegraphics[width=0.45\textwidth]{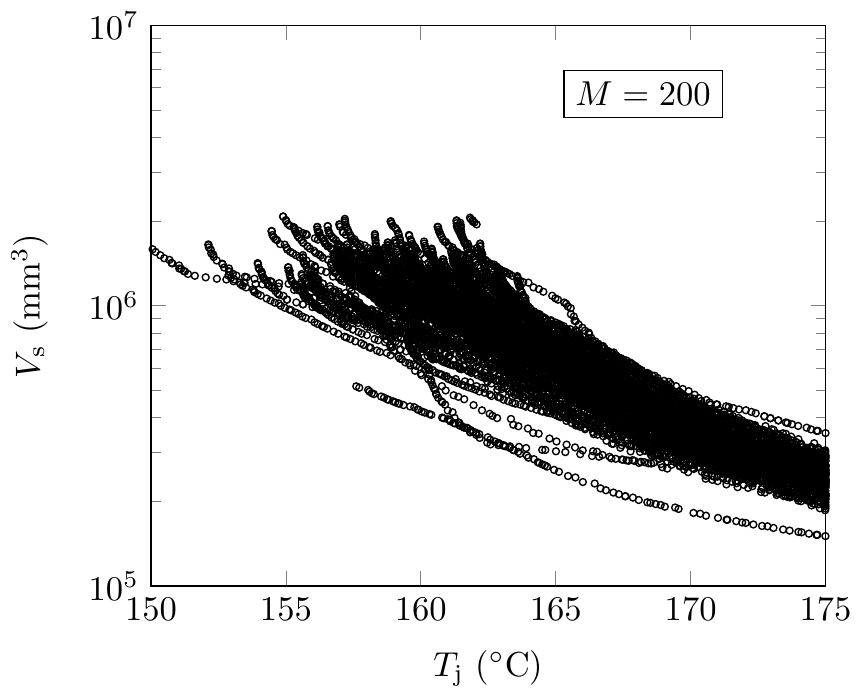}
	\\
	\includegraphics[width=0.45\textwidth]{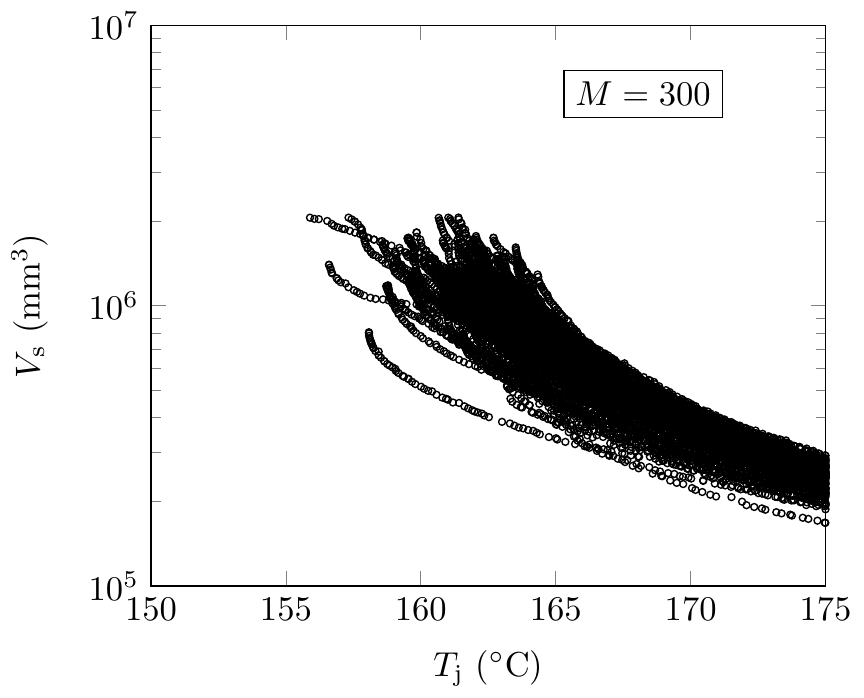}
	\hfill
	\includegraphics[width=0.45\textwidth]{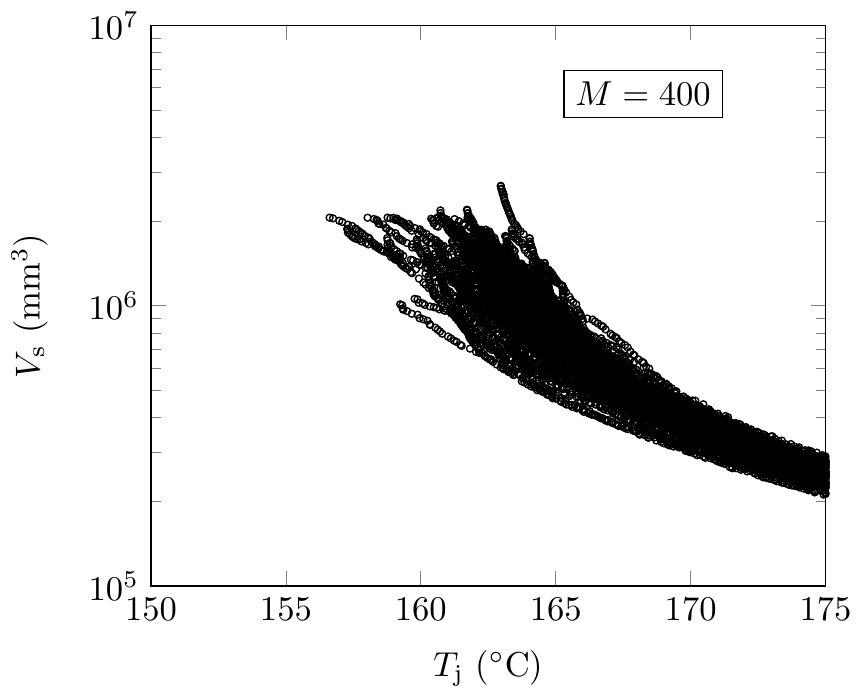}
	\\
	\includegraphics[width=0.45\textwidth]{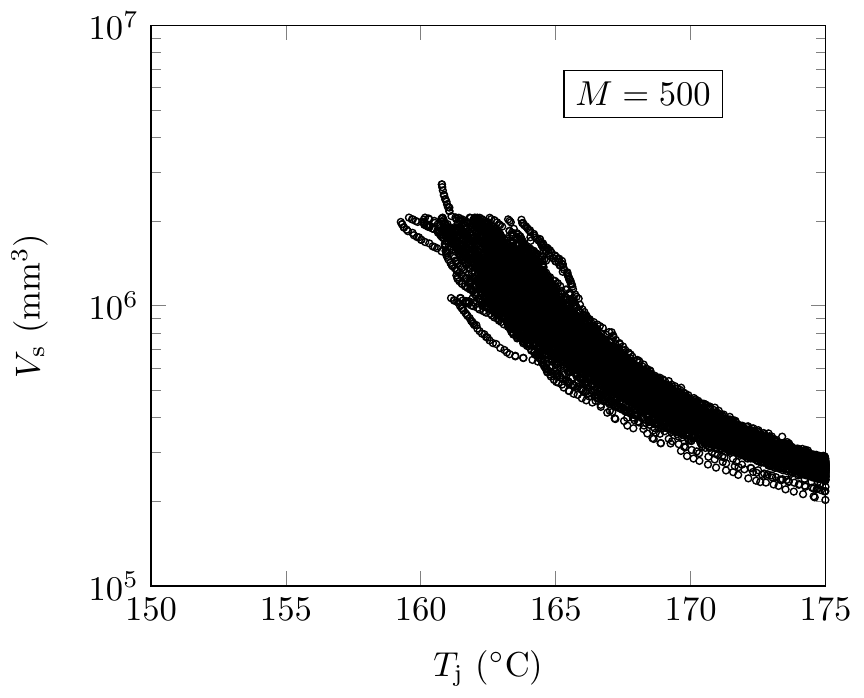}
	\hfill
	\includegraphics[width=0.45\textwidth]{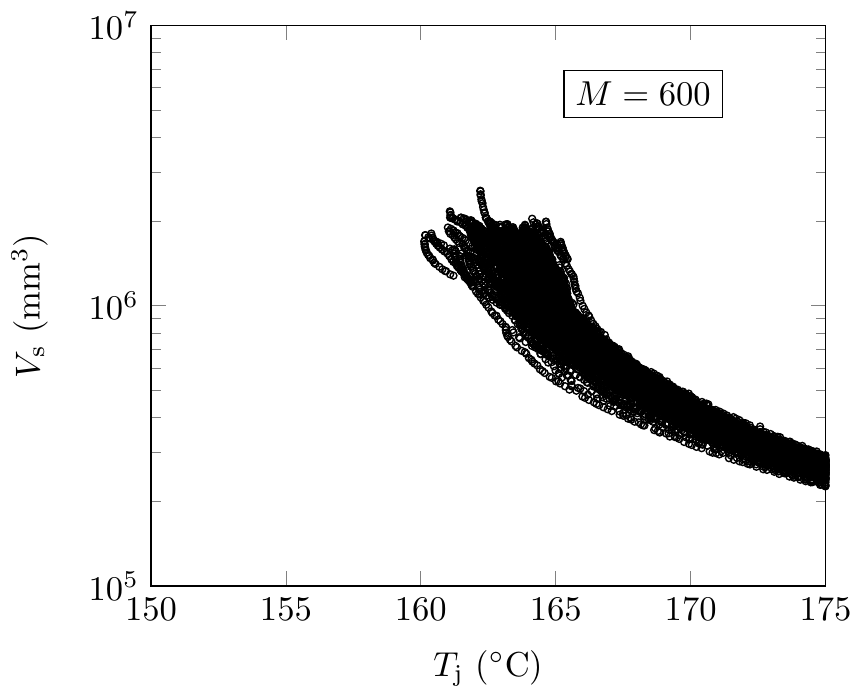}
	\\
	\includegraphics[width=0.45\textwidth]{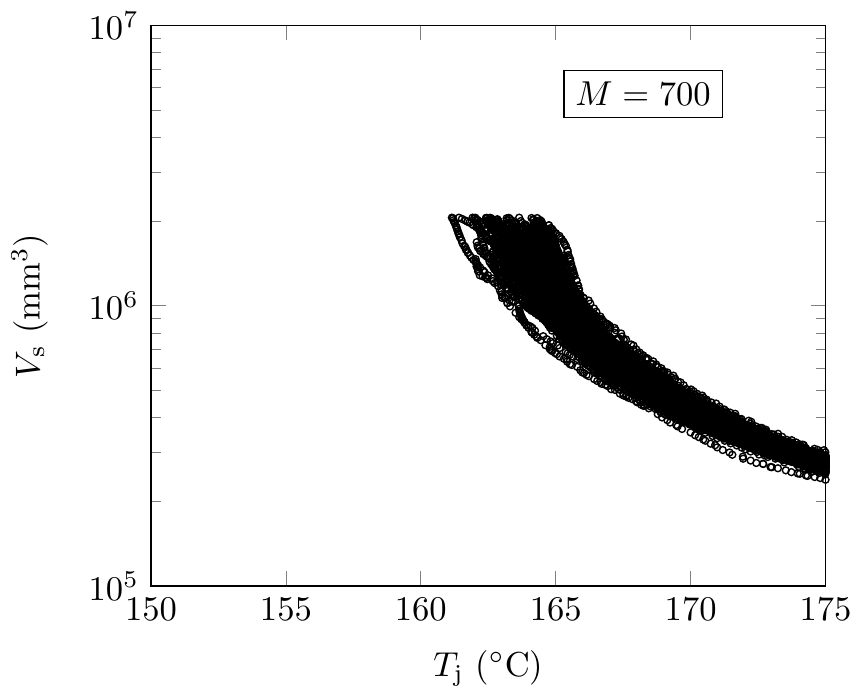}
	\hfill
	\includegraphics[width=0.45\textwidth]{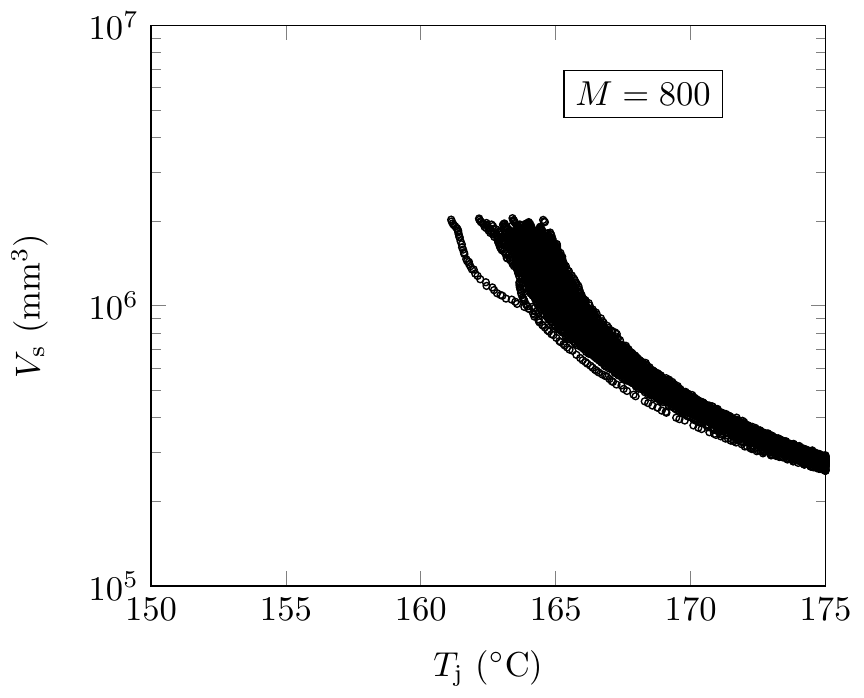}
	\caption{Pareto front solutions for each of the $I=100$ sparse-adaptive \gls{ddpce} surrogate models computed with training datasets of size $M=100,200,\dots,800$.}
	\label{fig:moo-results}
\end{figure}

Similar to Section~\ref{subsubsec:single-obj-optimization}, we again use the sparse-adaptive \gls{pce} models $\left\{\widetilde{S}^{(i)}_{M}(\mathbf{y})\right\}_{i=1}^{I=100}$ to compute $I$ Pareto fronts for each training dataset size $M=100,200,\dots,800$. 
Note that the surrogate model is employed to estimate the junction temperature $T_{\text{j}}$, while the volume of the heat sink is estimated using an analytical formula based on the geometrical parameters of the heat sink.
The corresponding results are presented in Figure~\ref{fig:moo-results}, where each panel corresponds to a different value for $M$ and all $I$ Pareto fronts are plotted within a panel.
Note that the Pareto fronts cannot be aggregated and averaged in a simple manner, e.g. as done with the single-objective optimization results presented in Tables~\ref{tab:single-opt-pso-results} and \ref{tab:single-opt-DE-results}, as each Pareto front comprises significantly different $T_{\text{j}}$-$V_{\text{s}}$ pairs, equivalently, a different Pareto front ``trajectory'' is obtained for each $i=1,\dots,100$. 
For the same reason, we do not provide optimal parameter values as in Tables~\ref{tab:single-opt-pso-results-params} and \ref{tab:single-opt-de-results-params}. 
The task of choosing an optimal parameter configuration given the Pareto front solutions is typically left to the expertise of the design engineer.

As expected, increasing the volume of the heat sink results in lower junction temperatures and this trend is captured by all surrogate models.
Nevertheless, it is now even more obvious that surrogate model accuracy plays a crucial role in the robustness of the optimization results, as can be observed from Figure~\ref{fig:moo-results}. 
In particular, the overall variation in the Pareto front solutions obtained with surrogate models of lower accuracy is very large, especially concerning heat sink design realizations corresponding to small junction temperature values. 
Increasing the accuracy of the surrogate models results in significantly more conservative Pareto fronts, which at the same time do not present as strong variations and the corresponding results can be considered to be robust. 
It is however still the case, that variations in the Pareto fronts are greater for the heat sink designs that correspond to the lowest junction temperature values, while a robustification of the Pareto fronts can be observed as the junction temperature increases.

Concerning the computational cost of the multi-objective optimization by means of the \gls{nsga} algorithm, the average number of surrogate model evaluations for each training dataset size $M=100,200,\dots,800$, is given in Table~\ref{tab:multi-opt-results}. 
In comparison to the single-objective optimization costs given in Tables~\ref{tab:single-opt-pso-results} and \ref{tab:single-opt-DE-results} for the \gls{pso} and \gls{de} algorithms, respectively, it is obvious that multi-objective optimization is significantly more expensive. 
In this case, using a surrogate model in the place of the original 3D \gls{cfd} model is even more advantageous in terms of computational demand.

\begin{table}[b!]
	\centering
	\caption{Average surrogate model evaluations for multi-objective optimization with the \gls{nsga} algorithm.}
	\begin{tabular}{|l|c|c|c|c|c|c|c|c|}
		\hline 
		Training dataset size $M$ & 100 & 200 & 300 & 400 & 500 & 600 & 700 & 800 \\ 
		\hline
		Average model evaluations & 13705 & 14305 & 14555 & 15080 & 14140 & 14480 & 13795 & 14560\\
		\hline
	\end{tabular}
	\label{tab:multi-opt-results}
\end{table}

\section{Conclusions}
\label{sec:concl}

This work presented a data-driven surrogate modeling framework based on the \glsfirst{ddpce} method, which was applied to alleviate the costs of single- and multi-objective heat sink design optimization. 
Of particular interest is the performance of the data-driven surrogate modeling approach in the context of small data learning, as the employed training datasets have relatively small sizes of $M=100,200,\dots,800$ data points. 
To avoid pitfalls in the small-data learning regime related to partitioning datasets to training and test data, an ensemble modeling technique was applied, which enhances the surrogate model's predictions with statistical information such as mean and variance, thus allowing to quantify model-form uncertainties arising due to limited data and assess the surrogate model in terms of robustness.

Based on the numerical results presented in Section~\ref{subsec:predictive}, surrogate models based on \glspl{ddpce}, in particular based on sparse-adaptive \glspl{pce}, are significantly more accurate and robust compared to competitive surrogate modeling methods based on \glspl{gp} and \glspl{nn}, at least for the specific test case considered in this work. 
Ensembles of sparse-adaptive \glspl{ddpce} were then used within stochastic single- and multi-objective optimization algorithms in Section~\ref{subsec:surrogate-based-optimization}, with the goal to identify suitable heat sink designs.
Therein, it is observed that the surrogate model plays a crucial role not only in the accuracy, but also in the robustness of the optimization results.
We note that the issues of model-form uncertainty and model robustness in surrogate-based design optimization studies have not received sufficient attention so far. The numerical results of this work show clearly that these are issues of importance that must be taken into consideration.
In terms of computational cost, surrogate-based optimization yields tremendous savings compared to the case where the original heat sink model would be used within the optimization algorithms. 


While already seemingly successful, a number of improvements to the \gls{ddpce}-based surrogate modeling framework considered in this work can be pursued in follow-up works.
Exemplarily, active learning techniques \cite{marelli2018active} can be employed to create more informative training datasets of reduced size, thus leading to improved surrogate model accuracy at comparatively lower computational cost.
Further improvement could be accomplished by using multi-element \gls{pce} methods \cite{chouvion2016development} to accommodate cases where the functional relation between design features and \gls{qoi} is not sufficiently smooth for a global polynomial approximation. 
The use of multi-fidelity surrogate modeling \cite{forrester2007multi} could also accelerate approximation convergence and further reduce computational costs.
Last, it would be of interest to examine whether the proposed surrogate-based heat sink design optimization approach extends beyond power electronics cooling, e.g. for heat sinks designed for microprocessors, optoelectronic devices, or various types of integrated circuits.

\section*{Acknowledgments}
Both authors acknowledge the support of the German Research Foundation (DFG) via the grant TRR 361 (project number:  492661287) and of the Graduate School Computational Engineering within the Centre for Computational Engineering at the Technische Universität Darmstadt.
The authors would like to thank our former M.Sc. student, Robin Scheich, for generating the original dataset of design feature combinations and corresponding model evaluations.

\bibliography{ddpce4hs}

\end{document}